\documentclass[%
 reprint,
 amsmath,amssymb,
 aps,
]{revtex4-2}

\usepackage{graphicx}
\usepackage{dcolumn}
\usepackage[Symbolsmallscale]{upgreek}
\usepackage{comment}

\usepackage{tikz}
\usetikzlibrary{decorations.pathmorphing}

\usepackage{accents}
\usepackage{bm}
\usepackage{hyperref}
\usepackage[color,final]{showkeys}

\def\Xint#1{\mathchoice
   {\XXint\displaystyle\textstyle{#1}}%
   {\XXint\textstyle\scriptstyle{#1}}%
   {\XXint\scriptstyle\scriptscriptstyle{#1}}%
   {\XXint\scriptscriptstyle\scriptscriptstyle{#1}}%
   \!\int}
\def\XXint#1#2#3{{\setbox0=\hbox{$#1{#2#3}{\int}$}
     \vcenter{\hbox{$#2#3$}}\kern-.5\wd0}}

\def\dashint{\Xint-}

\newcommand{\wt}[1]{\widetilde{#1}}
\newcommand{\wh}[1]{\widehat{#1}}

\newcommand{\mc}[1]{\mathcal{#1}}

\newcommand{\oring}[1]{\accentset{\circ}{#1}}
\newcommand{\tl}{\tilde}
\renewcommand{\i}{\text{i}}
\newcommand{\e}{\text{e}}
\newcommand{\sgn}{\text{sgn}}
\newcommand{\ud}{\text{d}}

\newcommand{\Jsn}{\text{sn}}

\begin{document}


\title{2+1 dimensional gravity in AAdS spacetimes with spatial
  wormhole slices:\\ 
  Reduced phase space dynamics and the BTZ black hole}

\author{Anurag Kaushal}
\email{anuragkaushal314@gmail.com}
\author{Naveen S. Prabhakar}%
 \email{naveen.s.prabhakar@gmail.com}
 \author{Spenta R. Wadia}
 \email{spenta.wadia@icts.res.in}
 \affiliation{%
 International Centre for Theoretical Sciences-\\
 Tata Institute of Fundamental Research, Shivakote, Bengaluru 560089, India.
}%




\begin{abstract}
  We solve Einstein's equations with negative cosmological constant in
  $2+1$ dimensions in the Hamiltonian formulation. The spacetime has
  the topology of $\Sigma \times \mathbf{R}$ where $\mathbf{R}$
  corresponds to the time direction and $\Sigma$ is a cylinder
  $\mathbf{R} \times \mathbf{S}^1$ and the spacetime metric satisfies
  asymptotically AdS (AAdS) boundary conditions. 
  
  We address the question of gauge invariance by fixing the maximal slicing and spatial harmonic gauge conditions and demonstrate that there are no residual small diffeomorphisms in this gauge.
  We explicitly solve the Hamiltonian and momentum constraints, and
  the gauge conditions to obtain a two dimensional reduced phase
  space. For simplicity, and with the BTZ black hole in mind, we
  restrict the solution of the momentum constraints to be independent
  of $\mathbf{S}^1$.
  
  
  
  In AAdS spacetimes besides the standard Wheeler-deWitt equations
  there is a Schroedinger equation corresponding to the boundary ADM
  Hamiltonian. We express this Hamiltonian in terms of the reduced
  phase space variables and discuss its classical solutions and
  quantization. We exhibit the wave functions and a continuous
  positive energy spectrum. Each energy eigenvalue $E$ corresponds to
  a BTZ black hole of mass $M=E/2$. This identification is based on 
  the fact that the classical solution of the reduced phase space dynamics 
  gives rise to a spacetime that is related to the two-sided BTZ black hole 
  by a diffeomorphism.
    
  
  \end{abstract}

\maketitle

\tableofcontents

\section{Introduction}

It is well-known that the theory of gravitation in $2+1$ dimensions
described by Einstein's equations has no local degrees of freedom
\cite{Deser:1983tn, Deser:1983nh, Gott:1982qg, Giddings:1983es, Witten:1988hc}. In
particular, this implies that there is no Newton's law of gravitation
in the weak-field limit and no gravitational force. It has been shown
by Banados, Teitelboim and Zanelli \cite{Banados:1992wn} that black
hole solutions exist in $2+1$ dimensional Einstein gravity when the
cosmological constant is negative; however, the singularity in the
interior of the black hole is not a curvature singularity and simply
separates regions containing closed time curves. This feature
of the $2+1$ dimensional black hole may also be attributed to the
absence of a local gravitational force.

Though there are no local degrees of freedom, the topology of the
$2+1$ dimensional spacetime could give rise to global degrees of
freedom. In this paper, we restrict our attention to spacetimes of the
form $\Sigma \times \mathbf{R}$ where $\Sigma$ is a cylinder. For this
case, we systematically obtain an explicit description of these global
degrees of freedom and their dynamics in the Hamiltonian formulation
of Einstein's theory using Dirac's method for constrained Hamiltonian systems \cite{Dirac:1950pj, dirac2001lectures}. 

The Hamiltonian formulation of general relativity is applicable to
spacetimes of the form $\Sigma \times \mathbf{R}$; we review basic
aspects in Section \ref{prelims}. The phase space variables are the
spatial metric $g_{ij}$ on the two dimensional $\Sigma$ and its
conjugate momentum $\pi^{ij}$. In addition, there are two auxiliary
fields: the lapse $N$ and the shift $N^i$, which correspond to the
time-time component and time-space components of the spacetime
metric. In the Hamiltonian formulation, the six Einstein's equations
in $2+1$ dimensions separate nicely into three constraint equations
which do not involve time derivatives of $g_{ij}$ and $\pi^{ij}$, and
three dynamical equations which involve first order time derivatives
of $g_{ij}$ and $\pi^{ij}$.

The three constraints -- the Hamiltonian constraint
$\mc{H}_\perp \approx 0$ and the two momentum constraints
$\mc{H}_i \approx 0$, $i = 1,2$ -- generate the diffeomorphisms of the
$2+1$ dimensional spacetime. We solve the constraint equations by
choosing gauge conditions for each of the three equations. These are
the maximal slicing condition $g_{ij} \pi^{ij} = 0$ for the
Hamiltonian constraint, and the spatial harmonic condition
$\partial_i(\sqrt{g} g^{ij}) = 0$ for the momentum constraints. In
Section \ref{gaugesec}, we demonstrate that these fix small
diffeomorphisms completely without any residual freedom in the case
that the spatial slices are cylinders or wormholes of the form
$\mathbf{R} \times \mathbf{S}^1$.

On a given spatial slice with the topology of the cylinder, the
solutions $g_{ij}$ and $\pi^{ij}$ of the constraints and gauge
conditions are determined completely in terms of two numbers $m$ and
$T$, which constitute the reduced phase space degrees of freedom. The
number $m$ is constrained to be non-zero and represents the modulus of
constant negative curvature metrics on the cylinder with AAdS boundary
conditions, and the quantity
$p_m = - \frac{4 \uppi^2}{\kappa^2} \frac{T}{m^2}$ turns out to be the
variable conjugate to $m$. Thus, the reduced phase space 
is two-dimensional with coordinates $(m, p_m)$. We
describe the solutions of the constraint equations and the structure
of the reduced phase space in Section \ref{purereduced}.

The preservation of the maximal slicing and spatial harmonic gauge
conditions under Hamiltonian evolution imposes differential equations
on the lapse $N$ and shift $N^i$ auxiliary fields. These differential
equations are elliptic in nature and have unique solutions given
appropriate boundary conditions for $N$ and $N^i$. We solve these
equations explicitly in Section \ref{largediff} for a given spatial
slice in terms of the reduced phase space data $m$ and $T$.

So far, we have described the solutions of the constraint equations on
a given spatial slice. However, to assemble the spacetime solution of
Einstein equations, we have to describe the dynamics of the data
$g_{ij}$, $\pi^{ij}$, $N$ and $N^i$ on the spatial slices. The
dynamics of $g_{ij}$, $\pi^{ij}$, $N$ and $N^i$ are completely
determined by the dynamics of the reduced phase space variables $m$
and $p_m$. In turn, the dynamics in reduced phase space is governed by
the reduced Hamiltonian which is obtained by plugging in the solutions
of the constraints and gauge conditions into the expression of the
original Hamiltonian of general relativity. We describe the reduced
Hamiltonian and the classical solutions of the reduced Hamiltonian
equations of motion in Section \ref{redphaseeq}. We also show that
this reduced phase space is (classically) related by a canonical
transformation to the phase space variables for spherically symmetric
spacetimes that were discussed by Kuchar \cite{Kuchar:1994zk} some
time ago. 

As we note in Section \ref{prelims}, in AAdS spacetimes there is
a Schroedinger equation corresponding to the boundary Hamiltonian
besides the standard Wheeler-deWitt equations (see
eq. \eqref{bdrySch}). In the present case the time that enters the
Schroedinger equation is the one that parametrizes the maximal
slices. Following this we discuss the quantization of the classical
reduced phase space and solve the Schroedinger equation in Section
\ref{sec:QM}. The energy levels $E$ correspond to BTZ black holes of
mass $M=E/2$. This identification is based on the fact (as we shall
explicitly show in section \ref{BTZdiffsec}), that the classical
solution of the reduced phase space gives rise to a spacetime related
to the two-sided BTZ black hole by a diffeomorphism.


The geometry of the spacetime solutions obtained thus far turns out to
be that of the Kruskal extension of the non-rotating BTZ black hole in
$2+1$ dimensions \cite{Banados:1992wn, Banados:1992gq} with a particular value of the ADM mass
determined by the initial conditions for the reduced phase space
trajectories. In Section \ref{BTZdiffsec}, we explicitly obtain the
diffeomorphism from the solution obtained in the Hamiltonian approach
above to the Kruskal extension of the BTZ black hole.

The maximally sliced solution has many interesting features which we
elucidate in Section \ref{BTZdiffsec}, some of which are: (1) the
spatial slices cut across the event horizons of the BTZ black hole,
and (2) the maximally sliced solution avoids the past and future
singularities of the black hole completely. It must be emphasized that
the presence of the event horizons in the maximally sliced solutions
becomes obvious only after the diffeomorphism to the Kruskal extension
of the BTZ black hole. It is of interest to note that this fact seems
analogous to an observation made sometime ago in \cite{Dhar:1993zz} in
the context of two dimensional string theory and its description in
terms of fermions: the presence of the stringy black hole
\cite{Mandal:1991tz, Witten:1991yr} of the $c=1$ matrix model in the
fermionic description was apparent only after a non-trivial
transformation of the phase space variables in the fermionic theory.


We conclude with a discussion of some interesting questions and
directions for future research in Section \ref{discsec}. A set of
appendices supplement the results in the main text with details of
calculations.


\section{Preliminaries: \\The Hamiltonian Formulation}\label{prelims}

We describe the Hamiltonian formulation of general relativity in
$d + 1$ dimensions for spacetimes which are asymptotically AdS (AAdS).
The Hamiltonian formulation can be written for spacetimes which take
the form $\Sigma \times \mathbf{R}$ where $\mathbf{R}$ corresponds to
time and $\Sigma$ is a $d$ dimensional spatial manifold. The spacetime
metric can be written in the ADM form:
\begin{align}\label{ADMform}
  \ud s^2 =& -N(x,{t})^2 \ud {t}^2 \nonumber \\
  &+ g_{ij}(x, {t}) \big(\ud x^i + N^i(x,{t}) \ud {t}\big)\big(\ud x^j + N^j(x,{t}) \ud {t}\big)\ . 
\end{align}
The basic dynamical variables are the Riemannian metric $g_{ij}$ on a
constant ${t}$ spatial slice $\Sigma$, and the phase space is
described by $g_{ij}$ and its conjugate momentum $\pi^{ij}$. The
Poisson brackets are
\begin{align}
  \{g_{ij}(x), \pi^{kl}(x')\} &= \delta_{(i}^{(k}\delta_{j)}^{l)}\,\delta(x,x')\ .
\end{align}
The Hamiltonian is
\begin{equation}\label{grham}
  H[N, N_i] = \int_\Sigma \ud^d x\,(N \mc{H}_\perp + N_i \mc{H}^i) + H_\partial\ ,
\end{equation}
where
\begin{enumerate}
\item $\mc{H}_\perp$ and $\mc{H}^i$ are resp.~the Hamiltonian and
  momentum constraints
  \begin{align}\label{grconst}
    \mc{H}_\perp
    &= \frac{\kappa^2}{\sqrt{g}}\Big(\pi^{ij} \pi_{ij} - \tfrac{1}{d-1}\pi^2\Big) - \frac{1}{\kappa^2}\sqrt{g}(R - 2\Lambda)\ ,\nonumber\\
    \mc{H}^i &= -2 D_j \pi^{ij}\ .
  \end{align}
  Here, $\pi = g_{ij}\pi^{ij}$ is the trace of $\pi^{ij}$.

\item $N$, $N_i$ are the auxiliary lapse and shift fields, with the
  lapse $N$ being constrained to be nowhere zero on $\Sigma$ in the
  classical theory.
  
\item The constant $\kappa^2 = 16\uppi G_N$ is the gravitational
  coupling which is set to $1$ unless indicated otherwise, and
  $\Lambda = -d(d-1)/2\ell^2$ is the cosmological constant with $\ell$
  the AdS length scale.

\item $H_\partial$ contains boundary terms that must be included to make
  the functional partial derivatives of $H$ with respect to $g_{ij}$
  and $\pi^{ij}$ well-defined
  \cite{Regge:1974zd,Henneaux:1985tv,Brown:1986nw}. The structure of
  the boundary terms depend on the type of boundary conditions that
  one imposes on the fields\footnote{See, for instance, the exposition
    in \cite{Anderson:2010ph,Witten:2018lgb} for a discussion of the
    various possible boundary conditions in gravity and their
    virtues.}. When all fields satisfy Dirichlet boundary conditions,
  with the spacetime metric approaching the AdS$_{d+1}$ metric at the
  asymptotic boundaries, we have
  \begin{equation}\label{grbdry}
    H_\partial  = \oint_{\partial\Sigma} \ud^{d-1} x \left(- \frac{2}{\kappa^2}\, N\sqrt{\sigma} k + 2 \hat{r}_i N_j \pi^{ij}\right) + H_{\rm ct}\ ,
  \end{equation}
  where $\sigma$ is the determinant of the induced metric on
  $\partial\Sigma$, $\hat{r}^i$ is the outward pointing unit normal of
  the boundary $\partial\Sigma$ and $k = g^{ij} D_i \hat{r}_j$ is the
  trace of the extrinsic curvature of the boundary.
\end{enumerate}
The term $H_{\rm ct}$ contains counterterms which are required to
subtract the divergences that occur due to the non-compactness of the
AAdS spacetimes \cite{Brown:1992br, Hawking:1995fd,
  Balasubramanian:1999re, Emparan:1999pm, Lau:1999dp, Mann:1999pc,
  Kraus:1999di}. In the $d = 2$ case that is studied in this paper,
the counterterm turns out be of the form
  \begin{equation}
d = 2:\quad    H_{\rm ct} = \frac{2}{\kappa^2} \oint_{\partial\Sigma} \ud x \, N\sqrt{\sigma} \frac{1}{\ell}\ .  
  \end{equation}

The lapse and shift are Lagrange multiplier fields and their equations
of motion give the Hamiltonian and momentum constraints
\begin{equation}
  \mc{H}_\perp \approx 0 \ ,\quad \mc{H}_i \approx 0\ ,
\end{equation}
where the $\approx$ symbol means we have to impose them weakly in phase
space, i.e., impose them only after computing any Poisson brackets
involving the expressions $\mc{H}_\perp $ and $\mc{H}_i$. These are
first class constraints since their Poisson brackets are again
proportional to a combination of these constraints. The algebra of the
constraints is \cite{Dirac:1951zz, Teitelboim:1972vw}
\begin{align}
  \{\mc{H}_\perp(x), \mc{H}_\perp(x')\} &= \big(g^{ij} \mc{H}_j(x) + g^{ij} \mc{H}_j(x')\big) \delta_{,i} (x, x')\ ,\nonumber\\
  \{\mc{H}_i(x), \mc{H}_\perp(x')\} &=  \mc{H}_\perp(x) \delta_{,i} (x, x')\ ,\nonumber\\
  \{\mc{H}_i(x), \mc{H}_j(x')\} &=  \mc{H}_i(x)\delta_{,j}(x,x') +  \mc{H}_j(x) \delta_{,i} (x, x')\ .
\end{align}
The remaining Hamiltonian equations of motion are the dynamical
equations \footnote{These equations can also be derived from the solutions of the Einstein-Hamilton-Jacobi equation along paths of steepest descent together with a notion of time which emerges from the geometry of the configuration space via the length of these paths measured with respect to the deWitt metric \cite{Kaushal:2024xob}.}
\begin{align}
  &\dot{g}_{ab} = 2N g^{-1/2} \left(\pi_{ab} - \tfrac{1}{d-1}g_{ab} \pi\right) + 2 D_{(a} N_{b)}\ ,\nonumber\\
  &\dot{\pi}^{ab}=\nonumber\\
  &\ \sqrt{g}\big(-N (R^{ab} - \tfrac{1}{2} g^{ab}R  +  \Lambda g^{ab}) + D^a D^b N - g^{ab} D^2 N\big)\nonumber\\
  &\  + \frac{N}{2\sqrt{g}} g^{ab}  (\pi^{ij} \pi_{ij} - \tfrac{1}{d-1}\pi^2) - \frac{2N}{\sqrt{g}} (\pi^{ac} \pi_c{}^b - \tfrac{1}{d-1}\pi \pi^{ab})\nonumber\\
  &\ -D_i(\pi^{ib} N^a) - D_j (\pi^{aj} N^b) + D_p(\pi^{ab} N^p)\ .
\end{align}
Plugging in the expression for $\pi^{ij}$ in terms of $\dot{g}_{ij}$
into the $\dot{\pi}^{ij}$ equation, we get the three second order
dynamical Einstein equations for $\ddot{g}_{ij}$.

The quantum theory relies on the ADM space+time split of the spacetime
metric \eqref{ADMform} which has constant ${t}$ slices as the manifold
$\Sigma$, and is described in terms of wavefunctionals of the spatial
metrics on the spatial slice $\Sigma$, $\Psi[g_{ij}]$. In the standard
way, the conjugate momentum $\pi^{ij}(x)$ acts on these
wavefunctionals as the functional derivative
$-\i\hbar \frac{\delta}{\delta g_{ij}(x)}$. The Schroedinger equation is
then
\begin{equation}\label{GRSchrodinger}
\i\hbar \frac{\partial}{\partial{t}} \Psi[g_{ij}, {t}] = \widehat{H} \Psi[g_{ij}, {t}]\ ,
\end{equation}
where ${t}$ is the time coordinate that appears in the ADM split
\eqref{ADMform}, and $\widehat{H}$ is the expression $H$ \eqref{grham}
with the fields $g_{ij}$, $\pi^{ij}$ promoted to operators and ordered
with a particular prescription. The first class constraints
$\mc{H}_\perp \approx 0$, $\mc{H}_i \approx 0$ become constraints on
the wavefunctionals \cite{DeWitt:1967yk}:
\begin{equation}\label{WDWeq}
  \widehat{\mc{H}}_\perp \Psi[g_{ij}, {t}] = 0\ ,\quad \widehat{\mc{H}}_i \Psi[g_{ij}, {t}] = 0\ .
\end{equation}
Plugging this back into the Schroedinger equation
\eqref{GRSchrodinger}, the volume integrals simply vanish and we are
left with the boundary terms in the Hamiltonian:
\begin{equation}\label{bdrySch}
\i\hbar \frac{\partial}{\partial {t}} \Psi[g_{ij}, {t}] = \widehat{H}_\partial \Psi[g_{ij}, {t}]\ .
\end{equation}
In this paper, we specialize to the case of $d = 2$, i.e., $2+1$
dimensional asymptotic spacetimes.  It is a known fact that Einstein
gravity in $2+1$ dimensions is non-dynamical, i.e., it does not
contain local dynamical degrees of freedom. However, there are usually
global degrees of freedom arising from the topology of the spacetime,
including boundaries. We shall see the consequences of this at various
stages of our analysis.

The Hamiltonian theory is analyzed by first solving the Hamiltonian
and momentum constraints by imposing the maximal slicing, spatial
harmonic gauge conditions, which we discuss in Section
\ref{gaugesec}. This way, we obtain a classical reduced phase space in
which the constraints are automatically satisfied. In the $2+1$
dimensional setting of the current paper, this reduced phase space
turns out to be finite dimensional and described by degrees of freedom
that refer to global topological aspects of the spacetime. Since the
constraints that appear in the volume integrals of the Hamiltonian
\eqref{grham} have been solved, the Hamiltonian for the reduced phase
space degrees of freedom simply becomes the boundary term $H_\partial$
evaluated on the reduced phase space.

\section{Asymptotic AdS boundary conditions}\label{AAdSbdry}

The asymptotic AdS boundary conditions that we specify make use of an
areal radial coordinate $r$ defined in a neighbourhood of a given
asymptotic boundary. The boundary is approached as $r \to
\infty$. This definition of AAdS boundary conditions is sufficient for
our purposes, though a coordinate independent version can be given
\footnote{see, for instance, \cite{Andersson:1996xd} and
  \cite{Hollands:2005wt} and references therein. For a definition in
  terms of the areal radial coordinate, see the classic treatments in
  \cite{Henneaux:1985tv, Brown:1986nw}.}. The global AdS$_{2+1}$
metric is
\begin{equation}
  \ud s^2 = -\left(\frac{r^2}{\ell^2} + 1\right) \ud t^2 + \left(\frac{r^2}{\ell^2} + 1\right)^{-1} \ud r^2 + r^2 \ud \varphi^2\ .
\end{equation}
The above metric has the following asymptotic ($r \to \infty$) values
for the lapse, shift, spatial metric and conjugate momenta:
\begin{align}\label{AdS3split}
  &\oring{g}_{ij} \ud x^i \ud x^j = \frac{\ell^2}{r^2} \ud r^2 + r^2 \ud \varphi^2\ ,\nonumber\\
  & \oring{\pi}^{ij} = 0\ ,\quad \oring{N} = \frac{r}{\ell}\ ,\quad \oring{N}_i = 0\ .
\end{align}
where we have used a circle over the symbols to indicate that they are
asymptotic values for the reference AdS$_{2+1}$ metric.

It then seems plausible to require that the variables $g_{ij}$,
$\pi^{ij}$, $N$ and $N_i$ of our solutions of Einstein's equations
approach the global asymptotic AdS values $\oring{g}_{ij}$,
$\oring{\pi}_{ij}$, $\oring{N}$, $\oring{N}_i$ respectively with some
\emph{a priori} unspecified fall-offs in $r$ as $r \to \infty$, i.e.,
as one approaches a given asymptotic boundary in the locally defined
areal radial coordinate (in case the solution is in a different
coordinate system, one has to first do a coordinate transformation to
an areal radial coordinate near the asymptotic boundary and then match
with the boundary conditions specified above).


However, we shall see that the spacetime solutions we obtain the
maximal slicing spatial harmonic gauge are such that the momentum
component $\pi^{rr}$ does not go to zero at the asymptotic boundaries
but to a time-dependent constant. Thus, we need to relax the above
requirement to allow for such boundary values of the momenta. The
consequences of relaxing the boundary conditions are interesting to
explore.

It turns out that the metric $g_{ij}$ indeed approaches the AdS values
in our solutions, and we continue with the requirement that $g_{ij}$
approach the asymptotic AdS value $\oring{g}_{ij}$. The following
consequence of this will be useful later. For solutions of the
Hamiltonian constraint $\mc{H}_\perp = 0$, the above boundary
conditions (including the relaxation on the boundary values of the
momenta) imply that the scalar curvature $R$ of $g_{ij}$ approaches
$2\Lambda$ as one approaches a given asymptotic boundary:
\begin{equation}
  R \to 2\Lambda = -\frac{2}{\ell^2}\ ,\quad\text{as}\quad r \to \infty\ .
\end{equation}

\section{Gauge Fixing}\label{gaugesec}

The constraints $\mc{H}_\perp \approx 0$ and $\mc{H}_i \approx 0$ are
first class constraints \footnote{They satisfy the \emph{surface
    deformation algebra} under the Poisson bracket \cite{Dirac:1951zz,
    dirac2001lectures, Teitelboim:1972vw}:
  \begin{align*}
  \{\mc{H}_\perp(x), \mc{H}_\perp(x')\} &= \big(g^{ij}\mc{H}_j(x) + g^{ij}\mc{H}_j(x')\big) \delta_{,i}(x,x')\ ,\\
  \{\mc{H}_i(x), \mc{H}_\perp(x')\} &=  \mc{H}_\perp(x) \delta_{,i}(x,x')\ ,\\
  \{\mc{H}_i(x), \mc{H}_j(x')\} &=  \mc{H}_i(x) \delta_{,j}(x,x') + \mc{H}_j(x) \delta_{,i}(x,x')\ ,
\end{align*}
where the derivatives on the $\delta$-functions are all with respect
to the first argument.} on phase space and generate gauge
transformations. These gauge transformations are generated with the
Hamiltonian \eqref{grham} with some choice for the auxiliary fields
$N$ and $N^i$, and they correspond to the diffeomorphisms of spacetime
which are trivial at the asymptotic boundaries -- the \emph{small
  diffeomorphisms}.

One fixes this gauge freedom by choosing gauge conditions, one for
each of the constraints. The solutions of the gauge conditions and
first class constraints describe the reduced phase space which contain
the truly gauge invariant degrees of freedom; an important requirement
is that gauge freedom is completely fixed without residual small
diffeomorphisms that preserve the gauge conditions. Note that large
diffeomorphisms (those diffeomorphisms that are non-trivial at the
boundaries but preserve the asymptotic AdS boundary conditions) need
not be affected by the gauge fixing procedure.

\subsection{The gauge conditions}

In this work, we choose the \emph{maximal slicing gauge} for the
Hamiltonian constraint $\mc{H}_\perp$ and the \emph{spatial harmonic
  gauge} for the momentum constraints $\mc{H}^i$:
\begin{align}\label{gaugecond}
  \text{Maximal slicing}:&\quad \pi = 0\ ,\nonumber\\
  \text{Spatial harmonic gauge}:&\quad \mc{D}^j := \partial_i(\sqrt{g} g^{ij}) = 0\ .
\end{align}
We show next that the maximal slicing gauge condition completely fixes
the small gauge transformations generated by the Hamiltonian
constraint $\mc{H}_\perp$ globally in phase space. Though we prove
this result in $d = 2$, it is readily generalized to $d > 2$ and we
emphasize that this result holds in $d > 2$ as well.

A similar general result does not hold for the spatial harmonic gauge
conditions. In this work, we specialize to $d = 2$ and spatial slices
with the topology of a cylinder
$\Sigma = \mathbf{R} \times \mathbf{S}^1$. In this case, we
demonstrate the spatial harmonic gauge conditions completely fix the
small gauge transformations generated by the momentum constraints
$\mc{H}_i$ globally in phase space. These are indeed the small spatial
diffeomorphisms on $\Sigma$, and imposing the harmonic gauge
conditions chooses a harmonic coordinate system on the spatial slice
$\Sigma = \mathbf{R} \times \mathbf{S}^1$.

However, we must note that the analysis for the spatial harmonic gauge
conditions typically fails for two dimensional spatial surfaces with
more complicated topology, and also in $d > 2$. That is, the spatial
harmonic gauge conditions cannot be imposed globally in phase
space. This is tied to the fact that spatial harmonic coordinate
systems typically fail beyond a certain region on the spatial slice
when the coordinate differentials become linearly
dependent.\footnote{We thank E.~Witten for pointing this out.} We note
that, in general, it is not known whether there exists a gauge
condition that holds globally in phase space for the momentum
constraints. \\

\subsection{The gauge fixing procedure}

To check that the gauge conditions completely fix the gauge
transformations generated by the Hamiltonian \eqref{grham}, we must
show that the Poisson brackets of the gauge conditions and the
constraints are not all (weakly) zero. It is straightforward to
compute the various Poisson brackets on the constrained surface
$\mc{H}_\perp \approx 0$, $\mc{H}_i \approx 0$, $\pi \approx 0$,
$\mc{D}^j \approx 0$:
\begin{widetext}
\begin{align}\label{pbconshamgauge}
  &\{ \mc{H}_\perp(x), \pi(x')\} \approx \kappa^{-2}{\sqrt{g}}\left(D^2  + 2\Lambda - \kappa^4 g^{-1}\pi_{ij} \pi^{ij}  \right) \delta(x,x')\ ,\quad \{ \mc{H}_i(x), \pi(x')\} \approx \pi(x') D_i \delta(x,x') \approx 0\ ,
\end{align}
\begin{align}\label{pbconsmomgauge}
  &\{\mc{H}_\perp(x), \mc{D}^j(x')\} \approx 2\kappa^2  \partial'_k\big(\pi^{jk}(x) \delta(x,x')\big)\ ,\quad \{\mc{H}^i(x), \mc{D}^j(x')\} \approx -2 D_k \partial'_l\big(\sqrt{g}(\tfrac{1}{2}(g^{ij} g^{kl} + g^{jk} g^{il} - g^{ik} g^{jl}) \delta(x,x'))\big)\ .
\end{align}
\end{widetext}
To verify that our gauge conditions completely fix the small
diffeomorphism gauge freedom, we must show that, given a spacetime
diffeomorphism $(N = \xi_\perp, N_i = \xi_i)$ that
\begin{itemize}
\item[(1)] is \emph{a small diffeomorphism}, i.e.,
  $\xi_\perp, \xi_i \to 0$ at the boundaries with sufficiently fast
  fall-offs, and
\item[(2)] preserves the gauge conditions \eqref{gaugecond}, i.e.,
  $\{H[\xi_\perp, \xi_i], \pi(x')\} = 0$,
  $\{H[\xi_\perp, \xi_i], \mc{D}^j(x')\} = 0$,
\end{itemize}
is the trivial diffeomorphism, i.e., $\xi_\perp = 0$, $\xi_i = 0$
everywhere on $\Sigma$. The condition
$\{H[\xi_\perp, \xi_i], \pi(x')\} = 0$ gives
\begin{equation}\label{xiperpdiff}
  \sqrt{g}\left(D^2  + 2\Lambda - \kappa^4\tfrac{1}{g}\pi_{ij} \pi^{ij} \right)\xi_\perp = 0\ .
\end{equation}
Multiplying by $\xi_\perp$, integrating over $\Sigma$, and integrating
by parts in the $D^2$ term, and discarding the boundary term since
$\xi_\perp$ is a small diffeomorphism \footnote{For spatial metrics
  $g_{ij}$ which approach the AdS$_d$ spatial metric
  $\frac{\ud r^2}{r^2} + r^2 \ud \Omega_{d-1}^2$, the fall-off
  condition on $\xi_\perp$ that ensures that the boundary contribution
  is zero is $\xi_\perp \sim r^{-(d-1)/2 -\epsilon}$.}, we get
\begin{align}\label{integral}
  &\int_\Sigma\ud^2 x\,\sqrt{g}\Big( D_i \xi_\perp D^i \xi_\perp + (\kappa^4 \tfrac{1}{g}\pi^{ij} \pi_{ij} -2 \Lambda) \xi_\perp^2\Big) = 0\ .
\end{align}
Every term on the left hand side is positive definite, and hence,
\eqref{integral} is satisfied only if $\xi_\perp = 0$. Thus, the only
solution to \eqref{xiperpdiff} is $\xi_\perp = 0$.

Next, the condition $\{H[\xi_\perp, \xi_i], \mc{D}^j(x')\} = 0$ gives
the differential equation
\begin{equation}\label{spadiff}
  \partial_l \big(   2\kappa^2 \xi_\perp \pi^{kl} + \sqrt{g}(2 {D}^{(k} \xi^{l)} -  {g}^{kl} {D}_i \xi^i )\big) = 0\ .
\end{equation}
Since we have already established that $\xi_\perp = 0$ previously, the
first term in the above equation drops out.

We now specialize to the spatial slice $\Sigma$ being a cylinder. In
Section \ref{purereduced} below, we will see that any metric $g_{ij}$
on the cylinder which is AAdS is conformally related to a flat
Cartesian metric $f_{ij}$ of the form
$\ud \theta^2 + m^2 \ud\varphi^2$. Note that the differential operator
in \eqref{spadifftl} is conformally invariant. Thus, the equation can
be written for the conformally flat Cartesian metric $f_{ij}$:
\begin{equation}\label{spadifftl}
  \partial_l \big(  \sqrt{f} (2 \partial^{(k} \xi^{l)} -  f^{kl} \partial_i \xi^i )\big) = 0\ .
\end{equation}
Contracting the above equation with $\xi_k$, integrating over
$\Sigma$,
integrating by parts and discarding the boundary term since $\xi_k$ is
a small diffeomorphism, we get
\begin{equation}\label{spadifftl3}
  \int_\Sigma \ud^2x\,  \sqrt{f} (\partial_{(k} \xi_{l)} - f_{kl} \partial\cdot \xi) (\partial^{(k} \xi^{l)} - f^{kl} \partial\cdot \xi) = 0\ .
\end{equation}
Since the integrand is positive definite, the above condition implies
the conformal Killing equation
\begin{equation}
	\partial_{(k} \xi_{l)} - f_{kl} \partial\cdot \xi = 0\ .
\end{equation}
Since the only small diffeomorphism which satisfies the conformal
Killing equation is the trivial diffeomorphism, we have shown that
$\{H[\xi_\perp, \xi_i], \mc{D}^j(x')\} = 0$ implies that $\xi_i = 0$
for the case of the cylinder
$\Sigma = \mathbf{S}^1 \times \mathbf{R}$.

One can now solve the constraints and gauge conditions simultaneously
to obtain the reduced phase space \footnote{The above analysis implies
  that the matrix of Poisson brackets
  $\mc{M}_{AB} = \{\mc{C}_A, \mc{C}_B\}$ is invertible, where
  $\mc{C}_A$ collectively denotes all constraints and gauge
  conditions. To analyze the Hamiltonian dynamics, one can
  alternatively work with the original phase but with the Poisson
  bracket replaced by the the Dirac bracket constructed out of the
  above matrix $\mc{M}_{AB}$ in the standard way.}.  The Poisson
bracket and reduced Hamiltonian on reduced phase space can be obtained
from the original Lagrangian $\pi^{ij} \dot{g}_{ij} - H$ by
substituting the solutions of the constraints and gauge
conditions.

In the next section, we describe the reduced phase space dynamics of
$2+1$ dimensional gravity with spatial slice
$\Sigma = \mathbf{S}^1 \times \mathbf{R}$, which satisfy the AAdS
boundary conditions discussed in Section \ref{AAdSbdry}. In Section
\ref{BTZdiffsec}, we show that the spacetime solutions that correspond
to the solutions in reduced phase space are diffeomorphic to the fully
extended non-rotating BTZ black hole \cite{Banados:1992wn,
  Banados:1992gq}.

\subsection{A note on the York gauge}

We would like to point out that the above discussion regarding
complete fixing of small diffeomorphisms also applies to the
\emph{York gauge}, also called the \emph{constant mean curvature}
(CMC) gauge,
\begin{equation}
  \text{York gauge}:\quad \pi(x) = \kappa^{-2} {g(x)}^{1/2}\ \tau\ ,
\end{equation}
where $\tau$ is a constant on the two dimensional spatial slice
$\Sigma$. The Poisson brackets \eqref{pbconshamgauge} of the gauge
condition $\pi - \kappa^{-2}g^{1/2} \tau$ with the Hamiltonian
constraint $\mc{H}_\perp$ and the momentum constraints $\mc{H}_i$ are
modified to
\begin{multline}\label{pbyork}
  \{ \mc{H}_\perp(x), \pi(x') - \kappa^{-2}g(x')^{1/2}\ \tau\} \\ \approx \kappa^{-2}{\sqrt{g}}\left(D^2  + 2\Lambda - \frac{\tau^2}{2} - \kappa^4 \frac{\hat\pi_{ij} \hat\pi^{ij}}{g}  \right) \delta(x,x')\ ,
\end{multline}
where $\hat{\pi}^{ij}$ is the traceless part of $\pi^{ij}$, and
\begin{multline}
  \{ \mc{H}_i(x), \pi(x')- \kappa^{-2} g(x')^{1/2}\ \tau\} \\ \approx \big(\pi(x') - \kappa^{-2}g(x')^{1/2}\ \tau\big) D_i \delta(x,x') \approx 0\ .
\end{multline}
When $\Lambda < 0$, which is the case in this paper, the differential
operator on the right hand side of \eqref{pbyork} is negative definite
since the total quantity $4\Lambda - \tau^2$ remains negative for all
values of $\tau$. This implies that there are no non-trivial small
diffeomorphisms $\xi_\perp$ which preserve the York gauge, exactly
analogous to the discussion for the maximal slicing gauge. In this
paper, we do not pursue the York gauge further and continue working
with the maximal slicing gauge condition.

When $\Lambda > 0$, which is the case with asymptotically de Sitter
spacetimes, the above differential operator is still negative definite
provided that the combination $4\Lambda - \tau^2$ is negative. This
places a restriction $\tau > \sqrt{4\Lambda}$ on the values of $\tau$
for which the York gauge completely fixes small diffeomorphisms. In
particular, this also means that the maximal slicing gauge condition
corresponding to $\tau = 0$ is not a good gauge condition when
$\Lambda > 0$. However, as we have seen, the York gauge can still be
employed in this situation.

The reduced phase for gravity in the York gauge in $d=2$ for compact
$\Sigma$ (which does not include the AAdS$_{2+1}$ case of the current
work) has been studied previously
\cite{Moncrief:1989dx,Andersson:1996dr}. Also see the recent paper
\cite{Chrusciel:2024vle} for general results on solving the constraint
equations in the AAdS$_{2+1}$ case.

\section{The reduced phase space}\label{purereduced}

We follow the Lichnerowicz-York method of decomposing the metric
$g_{ij}$ and conjugate momentum $\pi^{ij}$ as
\begin{equation}\label{Lichdecomp}
  g_{ij} = \e^{2\lambda} \tl{g}_{ij}\ ,\quad \pi^{ij} = \e^{-2\lambda} \tl{\pi}^{ij}\ .
\end{equation}
with the conformal factor $\lambda$ satisfying $\lambda \to 0$ as one
approaches the boundaries. We then solve the Hamiltonian and momentum
constraints subject to the maximal slicing and spatial harmonic gauge
conditions.

\subsection{The Hamiltonian constraint}

The condition $\lambda \to 0$ at the boundaries requires $\tl{g}_{ij}$
to satisfy the AAdS boundary conditions near the boundaries, in
particular, that its Ricci scalar $\tl{R}$ approach $2\Lambda < 0$
near the boundaries. However, it is always possible to Weyl transform
any such metric to one which satisfies $\tl{R} = 2\Lambda$
\emph{everywhere} on $\Sigma$ due to the existence and uniqueness of
solutions to Yamabe's equation.\footnote{Given any metric $g'_{ij}$
  whose Ricci scalar $R'$ approaches $2\Lambda$ near the boundary, we
  can always find a unique $\phi$ such that
  $\tl{g}_{ij} = \e^{2\phi} g'_{ij}$ has constant Ricci scalar
  curvature $2\Lambda$ everywhere. This $\phi$ is the unique
  solution of the Yamabe equation
  $D'^2 \phi = \tfrac{1}{2}R' - \Lambda \e^{2\phi}$. See
  \cite{Andersson:1992yk, Andersson:1996xd, Sakovich:2009nb,
    Witten:2022xxp, Wittentalk} for more details} Thus, we take
$\tl{g}_{ij}$ to satisfy $\tl{R} = 2\Lambda$ everywhere. Recall that
the maximal slicing condition is $\pi = 0$. This in turn imposes
$\tl{g}_{ij} \tl{\pi}^{ij} = \tl\pi = 0$.

Substituting the decomposition \eqref{Lichdecomp} in the Hamiltonian
constraint along with the maximal slicing condition, we get the
Lichnerowicz equation for $\lambda$:
\begin{equation}
2  \tl{D}^2 \lambda =  -\kappa^4  \tl{g}^{-1} \tl{g}_{ik} \tl{g}_{jl} \tl{\pi}^{ij} \tl{\pi}^{kl} \e^{-2\lambda} + 2\Lambda(1 - \e^{2\lambda})\ .
\end{equation}
It can be shown that, given a $\tl{g}_{ij}$ and $\tl\pi^{ij}$
satisfying $\tl{R} = 2\Lambda$ and $\tl\pi = 0$ everywhere on
$\Sigma$, the above Lichnerowicz equation has a unique solution
$\lambda$ which satisfies $\lambda \to 0$ as one approaches the AAdS
boundaries \cite{Andersson:1992yk, Andersson:1996xd, Sakovich:2009nb,
  Witten:2022xxp, Wittentalk}. We review the proof of this statement
in Appendix \ref{lichapp}.


In the case of $\Sigma = \mathbf{S}^1 \times \mathbf{R}$, the space of
constant negative curvature metrics which satisfy the AAdS boundary
conditions at both boundaries is very simple to describe. The spatial
harmonic gauge condition $\partial_i(\sqrt{g} g^{ij}) = 0$ is
conformally invariant so that $\tl{g}_{ij}$ must also satisfy it. It
turns out that any metric $\tl{g}_{ij}$ that satisfies
$\tl{R} = 2\Lambda$ everywhere on the cylinder and the spatial
harmonic gauge condition is of the form
\begin{equation}\label{constmet}
  \tl{g}_{ij} \ud x^i \ud x^j = \frac{\ell^2}{\cos^2\theta} (\ud \theta^2 + m^2 \ud\varphi^2)\ ,\quad\text{with}\quad m \neq 0\ .
\end{equation}
We outline the proof in Appendix \ref{hyperbolic}. Here,
$\theta \in [-\frac{\uppi}{2}, \frac{\uppi}{2}]$ is a longitudinal
coordinate with $\pm \uppi/2$ being the boundaries of the cylinder and
$\varphi \in [0,2\uppi]$ is an angular coordinate.

\emph{Thus, the space of metrics satisfying $\tl{R} = 2\Lambda$ on the
  cylinder and the spatial harmonic gauge condition is the space
  of the parameter $m\neq 0$ that appears in the
  metric \eqref{constmet}.}

\subsection{The momentum constraints}

It is convenient to express our equations in terms of the extrinsic
curvature
\begin{equation}
  \tl{K}^{ij} = \kappa^2\frac{1}{\sqrt{\tl{g}}} (\tl\pi^{ij} - \tl{g}^{ij}
\tl\pi)\ .
\end{equation}
The maximal slicing condition $\pi = 0$ translates to
$\tl{K} = \tl{g}_{ij} \tl{K}^{ij} = 0$. Making use of $ \tl{K} = 0$
and the harmonic coordinate system $(\theta,\varphi)$ introduced
above, the momentum constraints for pure gravity can be simplified to
\begin{equation}\label{coupledeq}
  \partial_\theta \tl{K}_{\theta\varphi} + \frac{1}{m^2} \partial_\varphi \tl{K}_{\varphi\varphi} = 0\ ,\quad \partial_\varphi \tl{K}_{\theta\varphi} - \partial_\theta \tl{K}_{\varphi\varphi} = 0\ .
\end{equation}
While it is possible to explicitly write down the most general
solutions to these equations, we restrict ourselves to solutions which
are independent of $\varphi$ since we are eventually interested in
maximally slicing the BTZ solution \footnote{The more general boundary
  conditions which include a $\varphi$ dependence are related to the
  Brown-Henneaux boundary gravitons.\cite{Brown:1986nw}. We could also
  take a non-zero constant value for $\tl{K}_{\theta\varphi} = J/2$,
  which would correspond to a rotating BTZ solution. We do not
  consider this generalization here though all our results can be
  extended to this case with minor modifications.}. Thus, the solution
is
\begin{equation}
\tl{K}_{\theta\theta} = \frac{T}{m^2}\ ,\quad \tl{K}_{\varphi\varphi} = -T\ ,\quad  \tl{K}_{\theta\varphi} = 0\ ,
\end{equation}
where $T$ is a real number independent of $\theta$ and $\varphi$. The
usual AAdS boundary conditions require $\tl{K}_{ij} = 0$ at the
boundaries since the AdS metric is static. However, to get non-trivial
dynamics, we need to relax the boundary conditions and allow for
non-zero boundary values of $\tl{K}_{ij}$.

The solutions to the momentum constraints are thus given by
\begin{align}\label{momconsol}
  &\tl\pi^{\theta\theta} = -\kappa^{-2}\frac{\cos^2\theta }{\ell^2} \frac{T}{m}\ ,\quad \tl\pi^{\varphi\varphi} = \kappa^{-2}\frac{\cos^2\theta }{\ell^2} \frac{T}{m^{3}}\ ,\nonumber\\
  &\tl\pi^{\theta\varphi} = 0\ .
\end{align}

\subsection{The Lichnerowicz equation}

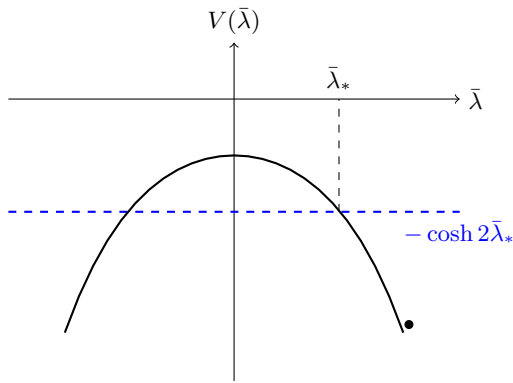
\begin{figure}
\begin{center}
  \begin{tikzpicture}[domain=-3:3,scale=0.75]
  \draw[->] (-4,0) -- (4,0) node[right] {$\bar\lambda$};
  \draw[->] (0,-5) -- (0,1) node[above] {$V(\bar\lambda)$};
  \draw[black,dashed] (1.86,-2) -- (1.86,0) node[above] {$\bar\lambda_*$};
  \draw[blue,dashed,thick] (-4,-2) -- (4,-2) node[below] {$-\cosh 2\bar\lambda_*$};
  \draw[color=black,thick] plot (\x,{-cosh(0.7*\x)});
  \draw [fill] (3.1,-4) circle [radius=2pt];
\end{tikzpicture}
\end{center}
\caption{The effective mechanics problem for the Lichnerowicz
  equation. The energy level $-\cosh 2\bar\lambda_*$ (blue dashed
  line) is such that the particle (black dot) stays on the positive
  side of the potential
  $V(\bar\lambda) = -\cosh 2\bar\lambda$.}\label{Lichparticle}
\end{figure}

\begin{figure}[!htbp]
  \centering
  \includegraphics[width=0.9\linewidth]{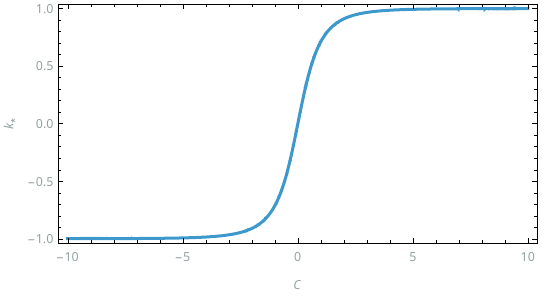}
  \caption{The function $k_*(C)$.}\label{kstarfn}
\end{figure}

Substituting the metric \eqref{constmet} and the solutions
\eqref{momconsol} into the Lichnerowicz equation, we get
\begin{equation}\label{Lichpde}
  \tl{D}^2 \lambda = -\frac{1}{\ell^2} - \frac{T^2 \cos^4\theta}{m^4 \ell^4} \e^{-2\lambda} + \frac{1}{\ell^2} \e^{2\lambda}\ .
\end{equation}
Since the coefficients of the above equation are independent of
$\varphi$, it can be truncated to the space of $\lambda$ which are
independent of $\varphi$.  Since there is a unique solution $\lambda$
for the above equation which satisfies $\lambda \to 0$ at the
boundaries \cite{Andersson:1992yk, Andersson:1996xd, Sakovich:2009nb,
  Witten:2022xxp, Wittentalk}, it must be the solution that is
independent of $\varphi$. Defining
\begin{equation}
  C = \frac{T}{\ell m^2}\ ,\quad   \bar\lambda(\theta) = \lambda(\theta) - \log \cos\theta - \tfrac{1}{2} \log |C|\ .
\end{equation}
we get the ordinary differential equation 
\begin{equation}
 \frac{1}{|C|} \bar\lambda '' = 2\sinh 2\bar\lambda\ ,
\end{equation}
which is the equation for a particle of mass $1/|C|$ with coordinate
$\bar\lambda$ and potential $-\cosh 2\bar\lambda$, with $\theta$ being
the time parameter ($' = \frac{\ud}{\ud\theta}$). Integrating the
equation above once, we get the `energy conservation' equation
\begin{equation}\label{particletraj}
  \frac{1}{2|C|}\bar\lambda'^2 - \cosh 2\bar\lambda = - \cosh 2\bar\lambda_*\ ,
\end{equation}
where $-\cosh 2\bar\lambda_*$ is the total `energy' of the particle
whereas the two terms on the left hand side are the kinetic and
potential energy respectively. The particle starts at `time'
$\theta = -\uppi/2$ at $\bar\lambda = \infty$, reaches the turning
point $\bar\lambda_*$ at $\theta = 0$ and returns to
$\bar\lambda = \infty$ at $\theta = \uppi/2$. See Figure
\ref{Lichparticle} for a sketch of the trajectory of the particle.
The energy conservation equation gives
\begin{equation}
  \bar\lambda' = \pm \sqrt{2|C|} \sqrt{\cosh 2\bar\lambda - \cosh 2\bar\lambda_*}\ ,
\end{equation}
corresponding to the $\theta > 0$ and $\theta < 0$ branches of the
trajectory respectively. The above first order equation can be solved
exactly and gives the following expression for the conformal factor
\begin{align}\label{lambdasol}
  \e^{\lambda(\theta)} &=  \sqrt{\frac{C}{k_*}}\, \frac{\cos\theta}{\Jsn\Big(\sqrt{\frac{C}{k_*}}\left(\tfrac{\pi}{2} - \theta\right); |k_*|\Big)}\ ,
\end{align}
where $\Jsn(x; k_*)$ is Jacobi's sn function, and is a function of $C$
determined by the transcendental equation
\begin{equation}\label{kstardef}
  \frac{\pi^2 C}{4} = k_* K(|k_*|)^2\ ,
\end{equation}
where $K(|k_*|)$ is Jacobi's complete elliptic integral of the first
kind. The function $k_*(C)$ satisfies
\begin{align}
  \sgn(k_*) &= \sgn(C)\ ,\nonumber\\
  k_*(C) &= C + \mc{O}(C^2)\quad\text{for}\quad C \approx 0\ ,\nonumber\\
  k_*(C) &\sim \pm(1 - 8\e^{-\pi\sqrt{|C|}})\quad\text{as}\quad C \to \pm\infty\ ,
\end{align}
and is numerically plotted in Figure \ref{kstarfn}. 

\begin{figure}[!htbp]
  \centering
  \includegraphics[width=0.9\linewidth]{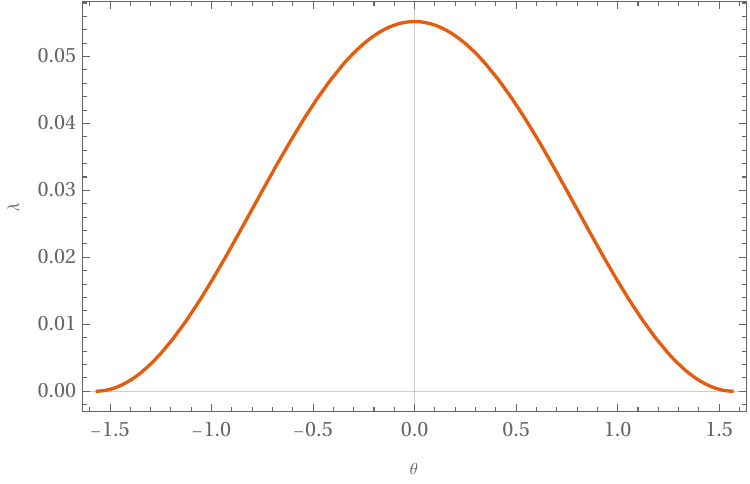}
  \caption{$\lambda(\theta)$ for $C = 0.5$, $k_* = 0.448$.}\label{lambdaplot}
\end{figure}

Substituting $\theta = 0$ in \eqref{lambdasol}, using \eqref{kstardef}
and $\Jsn\big(K(|k_*|); |k_*|\big) = 1$, we see that $k_*(C)$ is
related to the turning point value $\bar\lambda_*$ of $\bar\lambda$:
\begin{equation}
  |k_*| = \e^{-2\bar\lambda_*}\ .
\end{equation}
The solution \eqref{lambdasol} is a smooth, even function of $\theta$
which is well-defined everywhere in the range
$\theta \in [-\frac{\uppi}{2}, \frac{\uppi}{2}]$ and satisfies the
boundary conditions $\lambda(\pm \frac{\uppi}{2}) = 0$. See Figure
\ref{lambdaplot} for a plot of $\lambda(\theta)$ for the value
$C = 0.5$.
The solutions of the constraint equations and gauge conditions are
thus parametrized by two constants $m$ and $T$ which form the reduced
phase space. The final form of the solutions are
\begin{equation}\label{gfinal}
  g_{\theta\theta} = \e^{2\lambda(\theta)} \frac{\ell^2}{\cos^2\theta}\ ,\quad   g_{\varphi\varphi} = \e^{2\lambda(\theta)} \frac{\ell^2 m^2}{\cos^2\theta}\ ,\quad g_{\theta\varphi} = 0\ .
\end{equation}
\begin{align}\label{pifinal}
  &\pi^{\theta\theta} = \kappa^{-2}\e^{-2\lambda(\theta)} \frac{\cos^2\theta}{\ell^2}\frac{T}{m}\ ,\quad \pi^{\theta\varphi} = 0\  ,\nonumber\\
  &\pi^{\varphi\varphi} = -\kappa^{-2}\e^{-2\lambda(\theta)} \frac{\cos^2\theta}{\ell^2}\frac{T}{m^{3}}\ .
\end{align}

\subsection{The description of the reduced phase
  space}\label{redphdescr}

We see that, in the maximal slicing spatial harmonic gauge, the
spatial metric $g_{ij}$ in given entirely in terms of the single real
parameter $m$. The cotangent direction $p_m$ in $\pi^{ij}$ that
corresponds to changes in $m$ can be obtained as follows:
\begin{equation}
  p_m \equiv \int_\Sigma \ud^2x\, \pi^{ij} \frac{\partial g_{ij}}{\partial m}\ .
\end{equation}
Plugging in the expressions in \eqref{gfinal} and \eqref{pifinal}, we get
\begin{align}\label{pmdef}
  p_m 
      &= -\frac{4\uppi^2}{\kappa^2} \frac{T}{m^{2}}\ .
\end{align}
The reduced phase space is thus
two-dimensional with the coordinates $(m, p_m)$. We shall see in
Section \ref{redphase} that the $\pi^{ij} \dot{g}_{ij}$ term in the
unreduced, original phase space action gives rise to the combination
$p_m \dot{m}$ when we substitute into it the expressions
\eqref{gfinal} and \eqref{pifinal}. From this, we can read off the
symplectic form on the reduced phase space to be
$\ud p_m \wedge \ud m$. 

\section{Large diffeomorphisms: \\the lapse and shift}\label{largediff}

So far, in the context of gauge fixing, we have been concerned with
the Hamiltonian function $H[N, N_i]$ \eqref{grham} for small
diffeomorphisms $(N = \xi_\perp, N_i = \xi_i)$. We have seen that the
gauge conditions \eqref{gaugecond} completely fix the small
diffeomorphism freedom. However, it is possible that there are
residual diffeomorphisms which preserve the gauge conditions that do
not fall-off to zero as one approaches the asymptotic AdS
boundaries. These are the \emph{large diffeomorphisms}. These can be
obtained by again looking at the differential equations
\eqref{xiperpdiff} and \eqref{spadiff} but dropping the condition that
the diffeomorphism $(N, N_i)$ falls off to zero at the asymptotic
boundaries. The equations are
\begin{align}\label{largediffeq}
  &\sqrt{g}\left(D^2 - K_{ij} K^{ij}  + 2\Lambda\right)N = 0\ ,\nonumber\\
  &\partial_l \big(  \sqrt{g} (2 N K^{kl} + 2 {D}^{(k} N^{l)} -  {g}^{kl} {D}_i N^i )\big) = 0\ ,
\end{align}
where recall that $K^{ij}$ is the extrinsic curvature defined in terms
of the conjugate momentum as
\begin{equation}\label{Kijdef}
  K^{ij} = \kappa^2 \frac{1}{\sqrt{g}} (\pi^{ij} - g^{ij} \pi)\ .
\end{equation}
We assume that the $N$ and $N^i$ are independent of $\varphi$; these
are the simplest solutions to work with, and eventually we shall see
that the spacetime with this lapse and shift corresponds to the
maximally sliced, non-rotating BTZ solution. The more general
equations are of interest for uncovering the boundary Virasoro
asymptotic symmetries that are present in $2+1$ AAdS spacetimes
\cite{Brown:1986nw}. We do not consider them here, and address this
interesting problem in future work.

Plugging in the solutions of the constraint equations \eqref{gfinal},
\eqref{pifinal}, the differential equations become
\begin{align}
  &\frac{\ud^2 N }{\ud \theta^2} - (4 |C| \cosh 2\bar\lambda)N = 0\ ,\nonumber\\
  &\frac{\ud^2 N^\theta}{\ud \theta^2} = -2\sgn(T) \ell^{-1} \frac{\ud}{\ud\theta}(\e^{-2\bar\lambda} N)\ ,\quad \frac{\ud^2 N^\varphi}{\ud \theta^2} = 0\ .
\end{align}

\subsection{The solution for the lapse}
Consider the expression
\begin{equation}
  N(\theta) = \bar\lambda' {m} \left(c + b \int^\theta_{\theta_0}\frac{\ud\vartheta}{\bar\lambda'^2(\vartheta)}\right)\ ,
\end{equation}
with $\bar\lambda(\theta)$ the solution of the Lichnerowicz equation,
and $\theta_0$ some reference value on the spatial slice. It is
straightforward to check that this function satisfies the equation for
the lapse.

Recall that
\begin{equation}
  \bar\lambda' = \pm \sqrt{2|C|(\cosh 2\bar\lambda - \cosh
  2\bar\lambda_*)}\ ,
\end{equation}
with the sign $\pm$ corresponding to the ranges $\theta > 0$ and
$\theta < 0$ respectively. Thus, there are two branches for the
solution, one for each part of the spatial slice. We choose the lower
limit $\theta_0$ in the integral to be $\pm \uppi/2$ and the
integration constants $(c, b)$ and $(-\tl{c},-\tl{b})$ respectively
for the $\theta > 0$ and $\theta < 0$ branches. Thus, the lapse is
given by
\begin{equation}\label{lapsesol}
  N(\theta) = \left\{\def\arraystretch{2}\begin{array}{cc} |\bar\lambda'|{m} \left(c + b \int^\theta_{\frac{\uppi}{2}} \frac{\ud\vartheta}{\bar\lambda'^2(\vartheta)}\right) & \quad \theta > 0 \\  -|\bar\lambda'|{m} \left(-\tl{c} - \tl{b} \int^\theta_{-\frac{\uppi}{2}}\frac{\ud\vartheta}{\bar\lambda'^2(\vartheta)}\right) & \quad \theta < 0\ .\end{array}\right.
\end{equation}
Demanding smoothness of the solution at $\theta = 0$ by equating the
function values and derivatives at $\theta = 0$, we get the relations
\begin{equation}\label{bdef}
  b = \tl{b} = b_* \frac{\tl{c} + c}{2}\ ,
\end{equation}
where
\begin{align}\label{brel}
  b_* &=  -\frac{\left(\frac{C}{k_*}\right)^{3/2} (k_*^2-1)^2}{2E(|k_*|) + (k_*^2-1) K(|k_*|)} \ .
\end{align}

Thus, the solution for the lapse has two free parameters $c$ and
$\tl{c}$ which is appropriate for a solution of a second order
differential equation. The solution is an even function of $\theta$
when $\tl{c} = c$. \\ 

\noindent \textbf{The sign of the lapse:} Let us look at the equation
for the lapse:
\begin{equation}
  D^2 N = (K_{ij} K^{ij} -2 \Lambda) N\ .
\end{equation}
Suppose $c$ and $\tl{c}$ are both positive. In this case, since the
lapse is positive and grows in an unbounded manner at both the
boundaries, it must have at least one minimum in the interior. At any
minimum point, we have $D^2 N > 0$. The differential equation then
implies that the minimum value of the lapse must be positive since
$(K_{ij} K^{ij} -2 \Lambda) \geq 0$. At any maxima of $N$ in the
interior, the same argument says that the maximum values of $N$ must
be negative. This is clearly an impossible situation and hence there
is only one (global) minimum for $N$ in the interior at which it is
positive. It then follows that $N$ is positive everywhere on $\Sigma$
and never attains a zero. \\

\noindent \textbf{AAdS boundary conditions for the lapse:} In Section
\ref{arealsecDiff} below where we introduce an areal radial coordinate
$r$ on the spatial slice $\Sigma$, we show that the choice
$c = \tl{c} = 1$ corresponds to a lapse which approaches the
asymptotic AdS value $\frac{r}{\ell}$ in the limit $r \to \infty$ (see
Section \ref{AAdSbdry} for the boundary conditions).  With this choice
we get a foliation of the asymptotic AdS spacetime in which the bulk
time matches the boundary time. However, other values for $c$ or
$\tl{c}$ are still allowed by the formalism. In fact a general choice
of $c$ and $\tilde{c}$ gives vector fields that implements a more
general large diffeomorphism of the spacetime foliation. With this
understanding, we proceed with parameters $c$ and $\tl{c}$ being
arbitrary for now. We will put $c=\tilde{c}=1$ from Section
\ref{phasetraj} onwards.

\subsection{The solution for the shift}
The shift equations are much simpler to solve. The solution for
$N^\varphi$ is
\begin{equation}\label{shiftphi}
  N^\varphi(\theta) = (\tfrac{\pi}{2} - \theta) a_\varphi  +  ( \tfrac{\pi}{2} + \theta ) \tl{a}_\varphi\ ,
\end{equation}
for constants $a_\varphi$ and $\tl{a}_\varphi$. The solution for
$N^\theta$ is
\begin{equation}\label{shifttheta}
  N^\theta =  (\tfrac{\uppi}{2} - \theta)a_\theta +  ( \tfrac{\uppi}{2} + \theta )\tl{a}_\theta - 2\ell^{-1}\sgn(T) \int^\theta_0 \ud\vartheta\, \e^{-2\bar\lambda} N\ ,
\end{equation}
where $a_\theta$, $\tl{a}_\theta$ are constants and $N$ is the
solution \eqref{lapsesol}.

The simplest boundary conditions that one can impose on the shift are
to set them to be zero at the boundaries
$\theta = \pm \frac{\uppi}{2}$. In particular, this is justified for
the component $N^\theta$ since the $\theta$ coordinate takes values in
the finite $[-\uppi/2, \uppi/2]$, and a non-zero $N^\theta$ at the
boundaries will move the boundaries which is inconsistent.

Setting the shift to be zero at the boundaries determines the values
of the constants $a_\theta$, $a_\varphi$, $\tl{a}_\theta$,
$\tl{a}_\varphi$. For the choice $c = \tl{c} = 1$ on the lapse
\eqref{lapsesol} that implements the AAdS boundary conditions, we get
\begin{equation}
  a_\varphi = \tl{a}_\varphi = 0\ ,\quad a_\theta = -\tl{a}_\theta = -\frac{2}{\uppi\ell}\sgn(T) \int_0^{\uppi/2} \ud \vartheta\, \e^{-2\bar\lambda} N\ .
\end{equation}

\section{The Hamiltonian equations of motion in reduced phase space}\label{redphaseeq}

\subsection{The reduced Hamiltonian}


The reduced phase space consists of the metric and momentum that
satisfy the constraints and gauge conditions. The Hamiltonian
\eqref{grham} on such configurations reduces to the boundary term:
\begin{equation}\label{bdryham}
  H_\partial =  - \frac{2}{\kappa^2} \oint_{\partial\Sigma}\ud\varphi\, N\sqrt{\sigma} \left(k - \frac{1}{\ell}\right) + 2 \oint_{\partial\Sigma}\ud\varphi\, \hat{r}_i N_j \pi^{ij}\ ,
\end{equation}
The second term involving the shift does not contribute when the shift
satisfies the boundary conditions $N^i = 0$ on $\partial\Sigma$. On
the solutions \eqref{gfinal}, \eqref{pifinal} with the lapse
\eqref{lapsesol}, the first term evaluates to
\begin{align}\label{redpsham}
  H(m, T) &=  \frac{2\uppi (c + \tl{c})}{\kappa^2} \frac{T}{\ell}  \left( k_*(T /\ell m^2) + \frac{1}{k_*(T/\ell m^2)}\right) \ .
\end{align}

Next, we discuss the Hamiltonian equations of motion in reduced phase
space. Since the Hamiltonian depends only on the phase space variables
$T$ and $m$, and not on any time parameter, it will be conserved under
time evolution. Define
\begin{equation}\label{ADMmass}
  M =  {2\uppi} \frac{T}{\ell}  \left( k_*(T /\ell m^2) + \frac{1}{k_*(T/\ell m^2)}\right) \ .
\end{equation}
This conserved quantity is the `dimensionless mass', i.e., $M = \kappa^2 M_{\rm ADM}$, where $M_{\rm ADM}$ is the ADM mass of the solution measured at either asymptotic boundary. The ADM mass in asymptotically AdS$_3$ spacetime is defined as
\begin{equation}
	M_{\rm ADM} = -\frac{2}{\kappa^2}\oint \ud\varphi N \sqrt{\sigma} (k - \ell^{-1})\, .
\end{equation}

\subsection{The reduced phase space action and equations of
  motion}\label{redphase}

\begin{figure}[!htbp]
  \centering
  \includegraphics[width=0.9\linewidth]{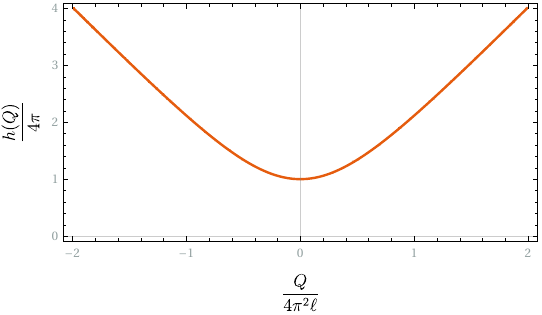}
  \caption{A plot of $\frac{1}{4\uppi} h(Q)$
    vs. $\frac{Q}{4\pi^2 \ell}$ for $c=\tl{c}=1$.}\label{hpmplot}
\end{figure}

The reduced phase space action is obtained from the action in the
original, unreduced phase space
\begin{equation}
  \mc{S} = \int \ud \bar{t} \int_\Sigma \ud^2x (\pi^{ij} \dot{g}_{ij} - H)\ ,
\end{equation}
where $H$ is the original Hamiltonian \eqref{grham}, and plugging in
the gauge-fixed expressions \eqref{gfinal} and \eqref{pifinal} for the
metric $g_{ij}$:
\begin{equation}\label{gfinal1}
  g_{\theta\theta} = \e^{2\lambda(\theta)} \frac{\ell^2}{\cos^2\theta}\ ,\quad   g_{\varphi\varphi} = \e^{2\lambda(\theta)} \frac{\ell^2 m^2}{\cos^2\theta}\ ,\quad g_{\theta\varphi} = 0\ ,
\end{equation}
and momentum $\pi^{ij}$:
\begin{align}\label{pifinal1}
  &\pi^{\theta\theta} = \kappa^{-2}\e^{-2\lambda(\theta)} \frac{\cos^2\theta}{\ell^2}\frac{T}{m}\ ,\quad \pi^{\theta\varphi} = 0\  ,\nonumber\\
  &\pi^{\varphi\varphi} = -\kappa^{-2}\e^{-2\lambda(\theta)} \frac{\cos^2\theta}{\ell^2}\frac{T}{m^{3}}\ .
\end{align}
The $\pi^{ij} \dot{g}_{ij}$ term in the action becomes
\begin{equation}
  \int_\Sigma\ud^2x\, \pi^{ij} \dot{g}_{ij} = p_m \dot{m} = -\frac{4\uppi^2}{\kappa^2} \frac{T \dot{m}}{m^{2}} \ .
\end{equation}
Let us define a new set of phase space coordinates,
\begin{equation}
	Q = 4\uppi^2\frac{T}{{m}^2}\ , \quad \text{and} \quad P = {m}\ ,
\end{equation}
and also define
\begin{equation}
	\bar{H} = \kappa^2 H \ .
\end{equation}
The reduced phase space equations of motion can then be derived from
the action
\begin{equation}\label{redaction}
  \mc{S}[Q, P] = \frac{1}{\kappa^2} \int \ud \bar{t} \left(P \dot{Q} - \bar{H}(Q, P)\right)\ ,
\end{equation}
where the Hamiltonian when expressed in terms of $Q$ and $P$ is
\begin{equation}
  	\bar{H}(Q,P) = P^2 h(Q)\ ,
\end{equation}
where
\begin{equation}
  h(Q) = \frac{c+\tl{c}}{2\uppi\ell} Q \left(k_*\left(\tfrac{Q}{4\uppi^2\ell}\right) + k_*\left(\tfrac{Q}{4\uppi^2\ell}\right)^{-1}\right)\ .
\end{equation} 
The function $h(Q)$ is plotted in Figure \ref{hpmplot}. 

The equations of motion are 
\begin{align}
  \dot{Q} = 2P h(Q)\ ,\quad \dot{P} = -(c+\tl{c}) \left(\frac{M}{Q} + \frac{\uppi^2 b_* P^2}{Q}\right)\ , 
\end{align}
where $b_*$ is the quantity given by \eqref{brel} which we reproduce
below for convenience:
\begin{equation}\label{brel1}
  b_* =  -\frac{\left(\frac{C}{k_*}\right)^{3/2} (k_*^2-1)^2}{2E(|k_*|) + (k_*^2-1) K(|k_*|)}\ .
\end{equation}
The equations for $m$ and $T$ can be obtained by using the definition
$Q = 4\uppi^2 \frac{T}{m^2}$: 
\begin{align}\label{mTeom}
  \dot{T} &=  -\frac{c+\tl{c}}{2}b_* m^{3}\ ,\nonumber\\
  \dot{m} &= -(c+\tl{c})\left(\frac{Mm^{2}}{4\uppi^2T} + \frac{b_* m^{4}}{4T}\right)\ . 
\end{align}


In Appendix \ref{Kuchar}, we show that our reduced phase space coordinates $(Q,P)$ are related by a canonical transformation to the phase space variables discussed by Kuchar \cite{Kuchar:1994zk} for spherically symmetric spacetimes applied to the BTZ black hole.\footnote{We thank Alok Laddha for bringing Kuchar's paper to our attention.}

\subsection{Phase space trajectories}\label{phasetraj}
From this section onwards we will consider AAdS boundary conditions which correspond to $c = \tl{c} = 1$.
As discussed above, the quantity $M$ \eqref{ADMmass} remains constant
under time evolution. Since the range of $k_*$ is $ 0 < |k_*| < 1$,
there is an upper bound for $|T|$ under time evolution which we call
$T_\infty$:
\begin{equation}
  |T| \leq T_\infty = \frac{M \ell}{4\uppi}\ .
\end{equation}
Further, the quantity $b_*$ is negative as can be seen from
\eqref{brel1}. The $\dot{T}$ equation of motion then implies that $T$
monotonically increases as $\bar{t}$ increases. Hence, as
$\bar{t} \to \pm\infty$, $T$ must approach $\pm T_\infty$. The
$\dot{m}$ equation of motion then implies that $m$ decreases
monotonically to $0$ as $\bar{t} \to \pm\infty$. In addition, the
above implies that $C = T / \ell m^2 \to \pm\infty$ so that
$k_*(C) \to \pm 1$. The conserved quantity $M$ then has a simple
expression in terms of $T_\infty$ which one obtains by taking
$\bar{t} \to \infty$:
\begin{equation}
  \frac{M}{2\uppi} = \frac{2 T_\infty}{\ell}\ .
\end{equation}
We shall see in Section \ref{BTZdiffsec} that the maximally sliced
solution is diffeomorphic to a region of the Kruskal extension of the
BTZ black hole so that ${M}$ corresponds to the mass of
the BTZ black hole. We have thus given an interpretation of the mass
of the BTZ black hole as the initial condition for one of the reduced
phase space variables.

\begin{figure}[!htbp]
  \centering
  \includegraphics[width=0.9\linewidth]{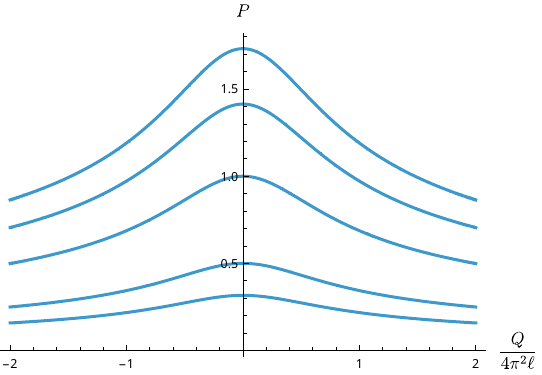}
  \caption{Phase space trajectories corresponding to
    $\frac{M}{2\uppi} = 3, 2, 1, \frac{1}{4}, \frac{1}{10}$ from top
    to bottom. Though not evident from the plots, all trajectories
    satisfy $P \to 0$ as $Q \to \pm \infty$. The value at $Q = 0$
    is $P=m= \sqrt{M / 2\uppi}$.}\label{phasetrajplot}
\end{figure}

We can solve the equations of motion explicitly as follows. Note that
the expression \eqref{ADMmass} for $M$ gives a quadratic equation for
$k_*$ whose solution gives an expression in terms of $M$ and $T$. This
is given by
\begin{equation}
  k_*(T) = \frac{R_-^2(T)}{\ell T} = \frac{\ell T}{R_+^2(T)} = \sgn(T)\frac{R_-(T)}{R_+(T)}\ ,
\end{equation}
where
\begin{equation}
  R_\pm^2(T) \equiv \ell T_\infty \left(1 \pm \sqrt{1 - \frac{T^2}{T_\infty^2}}\right)\ .
\end{equation}
Using $C = T / \ell m^2$ and the above expression for $k_*(T)$, we get
\begin{equation}\label{Tdoteq}
  \dot{T} = \frac{R_+}{\ell^3} \frac{(R_+^2 - R_-^2)^2}{2 R_+^2 E(\frac{R_-}{R_+}) + (R_-^2 - R_+^2) K(\frac{R_-}{R_+})}\ ,
\end{equation}
where we have suppressed the dependence of $R_\pm$ on $T$ for
brevity. This is an equation for $T$ that involves only $T$ on the
right hand side (apart from the constant of motion $M$) which makes it
easier to integrate. One can then plug this solution into the equation
for $m$ in \eqref{mTeom} and integrate the $\dot{m}$ equation to
obtain $m$ as a function of $\bar{t}$. Some sample phase space
trajectories are plotted numerically in Figure \ref{phasetrajplot}.

Before moving on, we note that while the total area under a phase
space trajectory is divergent, one can still get a finite area by
integrating from $Q=0$ to $Q=Q_c$:
\begin{align}
	\int_{0}^{Q_c} P\ud Q = \int_{0}^{Q_c} \sqrt{\frac{\bar{H}}{h(Q)}} \ud Q = \sqrt{2M} \int_{0}^{Q_c} \frac{\ud Q}{\sqrt{h(Q)}}\ ,
\end{align}
where $Q_c$ is some cut-off. The particular phase space trajectory
under consideration is uniquely labelled by the mass $M$. We will
revisit this fact later in Section \ref{sec:QM}.

\subsection{The collapse of the lapse}

The large $\bar{t}$ behaviour of $T$ and $m$ has an important
consequence on the lapse $N(\bar{t},\theta)$ of our solution.  We can
estimate the behaviour of $T$ and $m$ for large $\bar{t}$. Plugging in
$T \approx T_\infty$ in the equation \eqref{Tdoteq}, we get
\begin{equation}
  \dot{T} \approx 4 \left(\frac{T_\infty}{\ell}\right)^{3/2} \left(1 - \frac{T}{T_\infty}\right)\ ,
\end{equation}
which we can integrate explicitly to get
\begin{equation}\label{Tbeh}
  T = T_\infty(1 - B \e^{-2\sqrt{2} \eta \bar{t}})\ ,
\end{equation}
where we have defined $\eta$
\begin{equation}
  \eta = \frac{1}{\ell} \left(\frac{2T_\infty}{\ell}\right)^{1/2} = \frac{1}{\ell}\sqrt{\frac{M}{2\uppi}}\ ,
\end{equation}
which will be identified with the surface gravity of the BTZ black
hole.  $B$ in \eqref{Tbeh} is a constant of integration which we
cannot estimate from the large $\bar{t}$ behaviour, and has to be
determined from an exact integration of the $\dot{T}$ equation
\eqref{Tdoteq} either numerically or by an analysis similar to the one
in \cite{Beig:1997fp} for the maximally sliced Schwarzschild black
hole \cite{Estabrook:1973ue}.

Further, as $\bar{t} \to +\infty$, since $m$ becomes very small and
$T \to T_\infty$, one can drop the $m^{4}$ term in the $\dot{m}$
equation of motion and get
\begin{equation}
  \dot{m} \approx - \frac{M m^{2}}{2\uppi^2 T_\infty} = -\frac{2}{\uppi\ell} m^{2}\ ,
\end{equation}
which gives
\begin{equation}
  m(\bar{t}) \sim \frac{\uppi\ell}{2\bar{t}}\quad\text{as}\quad \bar{t} \to \infty\ .
\end{equation}
The value of the lapse $N(\bar{t}, \theta)$ at the centre of the
wormhole throat $\theta = 0$ takes the following form:
\begin{align}\label{collapse}
  N(\theta = 0,\bar{t})
  &= \frac{\ell \dot{T}(\bar{t})}{2\sqrt{T_\infty^2 - T^2}}\nonumber\\
  &\approx \ell\eta \sqrt{B} \, \e^{-\sqrt{2} \eta \bar{t}} \ ,\quad\text{as}\quad \bar{t} \to \infty\ .
\end{align}

The lapse is driven to zero as
$\bar{t} \to \infty$, a phenomenon termed as the \emph{collapse of the
  lapse} \cite{Estabrook:1973ue}. In the context of the mapping of the
solution to the Kruskal extension of the BTZ black hole, this means
that the proper time as measured by $N \ud \bar{t}$ is driven to
zero. This implies that the maximally sliced spacetime obtained in
this section will be mapped to only a subregion of the Kruskal
extension of the BTZ black hole, as we demonstrate in Section
\ref{BTZdiffsec}.

\section{Quantization of the reduced phase space}\label{sec:QM}


Recall the reduced phase space action \eqref{redaction} written in
$(m,T)$ variables:
\begin{equation}
	\mc{S}[{m}, T] = \int \ud\bar{t} \left(-\frac{4\uppi^2}{\kappa^2} \frac{T}{{m}^{2}} \dot{{m}} - H({m}, T)\right)\ .
\end{equation}  
The first term in the action above can be interpreted as arising from
the following symplectic form on a phase space with coordinates
$(m, T)$:
\begin{equation}
	\frac{4\uppi^2}{\kappa^2} \frac{\ud T \wedge \ud m}{ m^2}\ ,
\end{equation}
which is the symplectic form on the hyperbolic upper half-plane. It is
interesting that the Hamiltonian formulation of gravity naturally
provides the reduced phase space to be the hyperbolic upper half-plane
with its natural symplectic form. The quantization of the reduced phase
space may be carried out via the method of coadoint orbits since the
upper half-plane is indeed a coadjoint orbit of
$\text{SL}(2,\mathbf{R})$. It would also be interesting to identify
the origin of this $\text{SL}(2,\mathbf{R})$.

Presently this approach seems difficult to carry out explicitly and we
opt for directly setting up the Schroedinger equation and solving it
in terms of the coordinates
\begin{equation}
	Q = 4\uppi^2\frac{T}{{m}^2}\ , \quad \text{and} \quad P = {m}\ ,
\end{equation}
introduced before. Henceforth we interpret $Q$ and $P$ as `position and momentum' respectively.
Recall the reduced phase space action
\begin{equation}
	\mc{S}[Q, P] = \frac{1}{\kappa^2} \int \ud \bar{t} \left(P \dot{Q} - \bar{H}(P, Q)\right)\ ,
\end{equation}  
where the classical Hamiltonian $\bar{H}(P, Q)$ takes the form of the
kinetic energy of a free particle on a one dimensional space with
inverse metric $h(Q)$:
\begin{equation}\label{HDEF}
	\bar{H}(P,Q) =  h(Q) P^2\ ,
\end{equation}
with
\begin{equation}\label{hdef}
	h(Q) = \frac{Q}{\uppi\ell}  \left(k_*\left(\textstyle\frac{Q}{4\uppi^2\ell}\right) + \frac{1}{k_*\left(\frac{Q}{4\uppi^2\ell}\right)}\right)\ . 
\end{equation}
We have put $c=\tl{c}=1$ in the above and will keep this choice henceforth.

\begin{figure}[!htbp]
	\centering
	\includegraphics[width=0.9\linewidth]{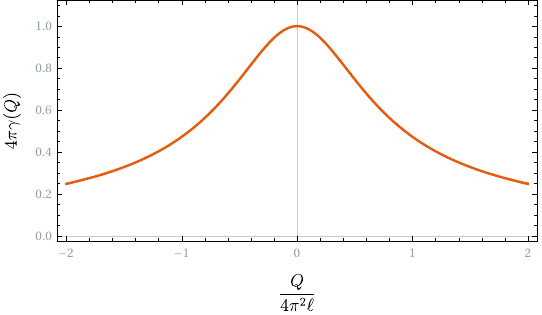}
	\caption{A plot of $4\pi\gamma(Q)$ vs. $\frac{Q}{4\pi^2 \ell}$.} \label{gammaplot}
\end{figure}

The quantization of a particle in curved space is well-known. A consistent choice of the quantum Hamiltonian is the Laplacian 
on the one dimensional manifold with metric $\gamma(Q) \equiv h^{-1}(Q)$:
\begin{align}
	\wh{H} &
	= - \hbar_{\text{eff}}^2 h^{1/2} \partial_Q \big(h^{1/2} \partial_Q \big) = -\hbar_{\text{eff}}^2 \frac{1}{\sqrt\gamma} \partial_Q \big(\frac{1}{\sqrt\gamma} \partial_Q \big)\ .
\end{align}
The metric $\gamma(Q)$ is plotted in Figure \ref{gammaplot}.

Note that in this problem $\hbar_{\text{eff}} \equiv \hbar\kappa^2$ 
is the effective Planck constant. The Schroedinger equation on the
position space wave functions $\psi(Q, \bar{t})$ is then
\begin{equation}
	\i \hbar_{\text{eff}} \frac{\partial \psi(Q, \bar{t})}{\partial \bar{t}} = - \hbar_{\text{eff}}^2 \frac{1}{\sqrt\gamma} \partial_Q \left(\frac{1}{\sqrt\gamma} \partial_Q \psi(Q, \bar{t})\right)\ .
\end{equation}
Since the Hamiltonian operator is time-independent, we can solve the
time-independent Schroedinger equation by separating variables
$\psi(Q, \bar{t}) = \e^{-\i E \bar{t}/\hbar_{\text{eff}}}\, \psi_E(Q)$:
\begin{equation}
	-\hbar_{\text{eff}}^2 \frac{1}{\sqrt\gamma} \partial_Q \left(\frac{1}{\sqrt\gamma} \partial_Q \psi_E(Q)\right) = E \psi_E(Q)\ .
\end{equation}
Let us define a new coordinate $\bar{Q}$ via
\begin{equation}
	\bar{Q}(Q) = \int_0^Q \ud  q \sqrt{\gamma(q)}\ .
\end{equation}
Since $\sqrt{\gamma(Q)} \sim Q^{-1/2}$ as $Q\to\infty$, the coordinate $\bar{Q}$ also has the same range $(-\infty, \infty)$ as $Q$. In terms of $\bar{Q}$, we get the simple equation
\begin{equation}
	-\hbar_{\text{eff}}^2 \frac{\partial^2}{\partial \bar{Q}^2} \psi_E(\bar{Q}) = E \psi_E(\bar{Q})\ ,
\end{equation}
whose spectrum is positive, i.e., $E \geq 0$, with the eigenfunctions
\begin{equation}
	\psi_p(\bar{Q}) = \e^{-\i p \bar{Q}/\hbar_{\text{eff}}}\ ,\quad\text{with}\quad p \in \mathbf{R}\ ,\ \text{and}\ E = p^2\ .
\end{equation}
These wave functions are $\delta$-function normalized:
\begin{equation}
	(\psi_p, \psi_{p'}) = \int_{-\infty}^\infty \ud\bar{Q}\, \psi_p^*(\bar{Q}) \psi_{p'}(\bar{Q}) = \delta(p - p')\ .
\end{equation}

The quantum mechanics of the reduced phase space in the $\bar{Q}$ variable is that of a free particle! The transition amplitude to go from $\bar{Q}_1$ to $\bar{Q}_2$ is well known
\begin{equation}\label{kernelQbar}
	K(\bar{Q}_2,\bar{t}_2;\bar{Q}_1,\bar{t}_1) = \frac{\exp\left(\frac{\i(\bar{Q}_2-\bar{Q}_1)^2}{4\hbar_{\text{eff}} (\bar{t}_2-\bar{t}_1)}\right)}{\sqrt{4\pi\i \hbar_{\text{eff}} (\bar{t}_2-\bar{t}_1) }}\ .
\end{equation}
The kernel in the `momentum' basis is the Fourier transform (here $\bar{Q} = \bar{Q}_2 - \bar{Q}_1)$
\begin{align}\label{tramp}
	K(p, \bar{t}_2-\bar{t}_1) &= \int_{-\infty}^{\infty} \ud \bar{Q} K(\bar{Q}_2,\bar{t}_2;\bar{Q}_1,\bar{t}_1) \e^{\i p \bar{Q}/\hbar_{\text{eff}}} \nonumber \\
	&= \exp\left(-\i\, p^2 (\bar{t}_2-\bar{t}_1)/\hbar_{\text{eff}} \right) \nonumber \\
	&= \exp\left(-\i\, 2M (\bar{t}_2-\bar{t}_1)/\hbar_{\text{eff}} \right)\ .
\end{align}

Since the Hamiltonian we have diagonalised is the boundary ADM Hamiltonian times $\kappa^2$, its eigenfunctions represent quantum states labeled by a real number $p$ with energy $E=p^2$. As we will discuss later in section \ref{BTZdiffsec}, the classical solutions in the reduced phase space correspond to the two sided BTZ black hole solution. Hence we can identify the energy eigenvalue with twice the mass of the BTZ black hole, i.e., $p^2=2M$. 

In the $Q$ representation the wave functions are
\begin{equation}
	\psi_p(Q) = \psi_p(\bar{Q}(Q))\ ,\quad\text{with}\quad \bar{Q}(Q) = \int_0^Q \ud  q \sqrt{\gamma(q)}\ ,
\end{equation}
and are normalized as:
\begin{equation}
	\int_{-\infty}^\infty \ud Q \sqrt{\gamma(Q)}\, \psi_p^*(Q) \psi_{p'}(Q) = \delta(p - p')\ .
\end{equation}
The transition amplitude in terms of $Q$ can be obtained from \eqref{kernelQbar}
\begin{align}
	K(Q_2,\bar{t}_2;Q_1,\bar{t}_1) &= \frac{\exp\left(\frac{\i \left(\int_{0}^{Q_2} \ud q\sqrt{\gamma(q)} -\int_{0}^{Q_1} \ud q\sqrt{\gamma(q)}\right)^2}{4\hbar_{\text{eff}} (\bar{t}_2-\bar{t}_1)}\right)}{\sqrt{4\pi\i \hbar_{\text{eff}} (\bar{t}_2-\bar{t}_1)}}\ .
\end{align}
The general wave function would be a linear combination of the above eigenstates,
\begin{equation}
	\psi(Q) = \int_{-\infty}^\infty \ud p\, C_p\, \psi_p(Q)\ .
\end{equation}
We require $C_{-p} = C^{*}_p$ for the wave function to be real.
Upon taking Fourier transforms
\begin{equation}
	\wt{\psi}(P) = \int_{-\infty}^{\infty} \ud Q\, \e^{-\i  Q P/\hbar_{\text{eff}}} \psi(Q)\ ,
\end{equation}
we obtain wave functions $\wt{\psi}(P)$ which are functions of $P$. 

Recall that the coordinate $P = m$ was classically restricted to be non-zero, however the quantum theory is completely well-defined for $P\in \bf{R}$. \\

\noindent \textbf{An exact classical-quantum correspondence:}
The phase of the energy eigenfunction with energy $E=2M$ is proportional to
\begin{equation}
	p \bar{Q} = p \int_{0}^{Q} \frac{\ud q}{\sqrt{h(q)}} = \sqrt{2M} \int_{0}^{Q} \frac{\ud q}{\sqrt{h(q)}}\ .
\end{equation}
Curiously this is the same as the `regulated' area under a phase space trajectory that we calculated in Section \ref{phasetraj},
\begin{align}
	\int_{0}^{Q_c} P\ud Q = \sqrt{2M} \int_{0}^{Q_c} \frac{\ud Q}{\sqrt{h(Q)}}\ ,
\end{align}
with the identification $Q=Q_c$.

Thus we see that each energy eigenstate labelled by $E=2M$ corresponds to a classical geometry which is diffeomorphic to the two-sided BTZ black hole of mass $M$ (as we show in the next section).

\section{The diffeomorphism to the extended BTZ black
  hole}\label{BTZdiffsec}

We have obtained a spacetime solution of Einstein's equations in the
Hamiltonian formulation in the the maximal slicing, spatial harmonic
gauge. The spacetime is the time evolution of a wormhole geometry with
two asymptotic AdS ends, with the spacetime metric in the ADM form by
\begin{align}\label{ADMmaximal}
  \ud s^2 &= -N(\bar{t},\theta)^2 \ud \bar{t}^2 + \frac{\ell^2 \e^{2\lambda(\bar{t},\theta)}}{\cos^2\theta} (\ud \theta^2 + N^\theta(\bar{t},\theta)\ud\bar{t})^2 \nonumber\\
  &\qquad\qquad\qquad\quad + \frac{\ell^2 \e^{2\lambda(\bar{t},\theta)} }{\cos^2\theta} m(\bar{t})^2 \ud\varphi^2\ .
\end{align}
This spacetime solution has ADM mass $\frac{M}{\kappa^2}$
\eqref{ADMmass} which is the value of the ADM Hamiltonian on the above
solution. It is straightforward to show that the above solution is in
fact diffeomorphic to a region $R_0$ of the full Kruskal extension of
the non-rotating BTZ black hole solution in $2+1$ dimensions. First,
we give a short review of the fully extended BTZ black hole solution.
In this section, we take $\kappa^2=1$ for convenience.

\subsection{A review of the fully extended BTZ black hole}\label{BTZrevmain}

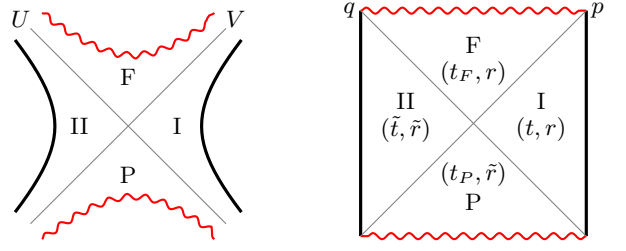
\begin{figure}
  \centering
  \begin{tikzpicture}[scale=0.65]
    \draw[black, very thick] plot[domain=-1.:1.] ({1.5*cosh(\x)},{1.5*sinh(\x)});
    \draw[black, very thick] plot[domain=-1.:1.] ({-1.5*cosh(\x)},{1.5*sinh(\x)});
    \draw[decorate, decoration={snake, amplitude=0.4mm, segment length=2.4mm}, thick, red] plot[domain=-1.:1.] ({1.5*sinh(\x)},{1.5*cosh(\x)});
    \draw[decorate, decoration={snake, amplitude=0.4mm, segment length=2.4mm}, thick, red] plot[domain=-1.:1.] ({1.5*sinh(\x)},{-1.5*cosh(\x)});
    \draw[gray] (-2,-2) -- (2,2);
    \draw[gray] (-2,2) -- (2,-2);
    \node[black] at (2.2,2.2) {$V$};
    \node[black] at (-2.2,2.2) {$U$};
    \node[black] at (1,0) {I};
    \node[black] at (0,1) {F};
    \node[black] at (-1,0) {II};
    \node[black] at (0,-1) {P};
  \end{tikzpicture}
\qquad\quad \begin{tikzpicture}[scale=0.75]
    \draw[black, very thick] (-2,-2) -- (-2,2);
    \draw[black, very thick] (2,-2) -- (2,2);
    \draw[decorate, decoration={snake, amplitude=0.4mm, segment length=2.4mm}, thick, red] (-2,2) -- (2,2);
    \draw[decorate, decoration={snake, amplitude=0.4mm, segment length=2.4mm}, thick, red] (2,-2) -- (-2,-2);
    \draw[gray] (-2,-2) -- (2,2);
    \draw[gray] (-2,2) -- (2,-2);
    \node[black] at (1.2,0.4) {I};
    \node[black] at (1.2,-0.1) {$(t,r)$};
    \node[black] at (0,1.4) {F};
    \node[black] at (0,0.9) {$(t_F,r)$};
    \node[black] at (-1.2,0.4) {II};
    \node[black] at (-1.2,-0.1) {$(\tl{t},\tl{r})$};
    \node[black] at (0,-1.4) {P};
    \node[black] at (0,-0.9) {$(t_P,\tl{r})$};
    \node[black] at (2.2,2) {$p$};
    \node[black] at (-2.2,2) {$q$};
  \end{tikzpicture}
  \caption{\label{BTZKruskal} The four regions of the maximally
    extended BTZ black hole shown in the Kruskal and Penrose
    diagrams. The local BTZ-type coordinates in each Region are shown
    in the Penrose diagram, with $r > R_h$ in Region I, $r < R_h$ in
    Region F, $\tl{r} > R_h$ in Region II and $\tl{r} < R_h$ in Region
    P.}
\end{figure}

The BTZ black hole metric is
\begin{equation}\label{BTZmetmain}
  \ud s^2 = - f(r) \ud t^2 + \frac{\ud r^2}{f(r)} + r^2 \ud\varphi^2\, ,\quad f(r) = \frac{r^2}{\ell^2} - \frac{M}{2\uppi}\ .
\end{equation}
The spacetime has an asymptotic AdS boundary as $r \to \infty$. The
metric has a coordinate singularity at the horizon
$r = R_h = \ell \sqrt{M / 2\uppi}$ at which the time translation
Killing vector becomes null.  The above metric is valid for the range
\begin{equation}
  \text{Region I}:\quad R_h < r < \infty\ ,\ - \infty < t < \infty\ ,\ 0 \leq \varphi < 2\uppi\ .
\end{equation}
To go past the null horizon, one first defines the tortoise coordinate
$r_*$:
\begin{equation}
  r_* = \int_\infty^r \frac{\ud \rho}{f(\rho)} = \frac{1}{2\eta} \log \frac{r - R_h}{r + R_h}\ ,
\end{equation}
where $\eta = R_h / \ell^2$ is the surface gravity of the black
hole. The Kruskal-Szekeres coordinates are defined as
\begin{align}\label{KruskalUV}
  U &= -\eta^{-1} \exp\big(-\eta(t - r_*)\big) = - \eta^{-1} \sqrt{\frac{r - R_h}{r + R_h}}\ \e^{-\eta t}\ ,\nonumber\\
  V &= \eta^{-1} \exp\big(\eta(t + r_*)\big) = \eta^{-1} \sqrt{\frac{r - R_h}{r + R_h}}\ \e^{\eta t}\ .
\end{align}
In Region I, we have $U < 0$ and $V > 0$. The past horizon
$t \to -\infty$, $r \to R_h$ is described by the surface $V = 0$
whereas the future horizon $t \to \infty$, $r \to R_h$ is described by
the surface $U = 0$. The BTZ black hole metric in terms of the
Kruskal-Szekeres coordinates is
\begin{equation}\label{BTZkruskal}
  \ud s^2 = -4 \frac{R^2_h}{\ell^2} \frac{\ud U \ud V}{(1 + \eta^2 UV)^2} + R_h^2 \left(\frac{1 - \eta^2 UV}{1 + \eta^2 UV}\right)^2 \ud \varphi^2\ .
\end{equation}
As is clear, the metric has no singularities at the horizons $UV = 0$,
and is valid for the range $|\eta^2 UV|<1$, with $U, V \in \mathbf{R}$.
This is the fully extended BTZ black
hole. There are four regions corresponding to the different signs of
$U$ and $V$ which we designate Region I, F, II and P. These are
displayed in Figure \eqref{BTZKruskal}.

In each of the regions, we can find BTZ-like coordinates in which the
metric takes the form \eqref{BTZmetmain}. These coordinates are also
displayed in Figure \eqref{BTZKruskal}. It is also useful to define
the Penrose coordinates $(p, q)$:
\begin{align}
  p = \arctan(\eta V)\ ,\quad q = \arctan(\eta U)\ .
\end{align}
The Penrose diagram of the fully extended BTZ black hole is displayed
in the second panel in Figure \ref{BTZKruskal}. The entire extended
black hole spacetime fits inside the finite region in $(p, q)$ space
where the $p$ and $q$ coordinates are in the range
$[-\frac{\uppi}{2}, \frac{\uppi}{2}]$. The singularities are at
$p + q = \pm \frac{\uppi}{2}$ and the boundaries at
$p - q = \pm \frac{\uppi}{2}$. The horizons are at $p = 0$ and
$q = 0$.

\subsection{The diffeomorphism} \label{arealsecDiff}

The strategy to obtain the diffeomorphism is as follows. We first
express our maximally sliced solution \eqref{ADMmaximal} in terms of
an areal radial coordinate since it facilitates an easy comparison
with the static BTZ metric in \eqref{BTZmetmain}.

\emph{Note that doing this takes us out of the spatial harmonic gauge
  with coordinates $(\theta, \varphi)$ whereas the maximal slicing
  gauge condition is still valid.}

The areal radial coordinate is defined by setting
$g_{\varphi\varphi}(\theta, \bar{t})$ in \eqref{ADMmaximal} equal to
$r^2$ and solving for $r$ in terms of $\theta$ and $\bar{t}$:
\begin{equation}\label{arealrad}
  r(\theta,\bar{t}) = \e^{\lambda(\theta,\bar{t})} \frac{\ell {m(\bar{t})}}{\cos\theta} = \e^{\bar\lambda(\theta,\bar{t})}  \sqrt{\ell|T(\bar{t})|}\ .
\end{equation}
where we have used the definition
$\bar\lambda = \lambda - \log \cos\theta - \frac{1}{2} \log
\frac{|T|}{\ell m^2}$. Recall that $\lambda$ is the unique solution to
the Lichnerowicz equation \eqref{Lichpde}, and it is an even function
of $\theta$. This implies that the above relation is a one-to-one
coordinate transformation either in the range $\theta > 0$ or in the
range $\theta < 0$. The range of $r$ that corresponds to
$0 < \theta < \frac{\pi}{2}$ is
\begin{equation}\label{arealrange}
R_+(\bar{t}) = \sqrt{\frac{\ell |T|}{|k_*|}} < r < \infty\ .
\end{equation}
Thus, we need two areal radial coordinates $r$ and $\tl{r}$ with
identical ranges to describe the entire spatial slice:
\begin{align}
  r &= \e^{\lambda(\theta)} \frac{\ell {m}}{\cos\theta} = \e^{\bar\lambda(\theta)}   \sqrt{\ell|T|}\ ,\quad\text{for}\quad \theta > 0\ ,\nonumber\\
  \tl{r} &= \e^{\lambda(\theta)} \frac{\ell {m}}{\cos\theta} = \e^{\bar\lambda(\theta)}  \sqrt{\ell|T|}\ ,\quad\text{for}\quad \theta < 0\ .
\end{align}
The maximally sliced spacetime metric in the coordinates
$(\bar{t}, r, \varphi)$ for the region $0 < \theta < \frac{\uppi}{2}$
is given by
\begin{align}\label{maxmetmain}
  &\ud s^2 =\nonumber\\
  & -N^2(r, \bar{t}) \ud \bar{t}^2 + A(r,\bar{t}) \left( \ud r + \frac{N(r,\bar{t}) T(\bar{t})}{r} \ud \bar{t}\right)^2 + r^2 \ud\varphi^2\ , \\
  & A(r,\bar{t}) =\left(\frac{r^2}{\ell^2} - \frac{M}{2\uppi} + \frac{T(\bar{t})^2}{r^2}\right)^{-1} \ ,
\end{align}
where the lapse $N(r,\bar{t})$ is given by
\begin{align}
  &N(r,\bar{t})\nonumber\\
  &\ = \frac{1}{\sqrt{A(r,\bar{t})}}  \Bigg(1 - \dot{T}(\bar{t}) \int_\infty^r \frac{\ud \rho}{\rho\left(\frac{\rho^2}{\ell^2} - \frac{M}{2\uppi} + \frac{T(\bar{t})^2}{\rho^2}\right)^{3/2}} \Bigg) ,
\end{align}
which is valid for the range \eqref{arealrange}.
The same expressions in terms of $\tl{r}$ hold for the region
$-\frac{\uppi}{2} < \theta < 0$ of the spatial slices.

Next, we postulate a diffeomorphism from Region I of the BTZ black
hole to the maximally sliced solution \eqref{maxmetmain} which maps
$t$ to a function of $\bar{t}$ and $r$:
\begin{equation}
(t, r,\varphi) = \big( t = t(\bar{t}, r), r, \varphi\big)\ .
\end{equation}
To obtain the function $t(\bar{t}, r)$, we plug it into the BTZ metric
\eqref{BTZmetmain} in Region I, and compare with the maximal slicing
metric \eqref{maxmetmain}. In Region I with $\bar{t} > 0$, we get
\begin{equation}
  \frac{\partial t}{\partial \bar{t}} = N(r, \bar{t}) A(r, \bar{t})^{1/2}\ ,\quad   \frac{\partial t}{\partial r} = -\frac{T(\bar{t}) A(r,\bar{t})^{1/2}}{r f(r)}\ .
\end{equation}
The solution to the above equations is obtained in Appendix \ref{BTZdiffapp}:
\begin{equation}
  t(r,\bar{t}) = \bar{t} + T(\bar{t}) \int_r^\infty \frac{\ud \rho}{\rho} \frac{A(\rho,\bar{t})^{1/2}}{f(\rho)}\ ,\quad r > R_h\ .
\end{equation}
Note that the integrand has a pole at $\rho = R_h$ coming from the
$f(\rho)$ in the denominator and hence $t \to \infty$ as $r \to
R_h$.

\emph{Thus, the diffeomorphism between the maximal slicing solution
  and the static BTZ coordinate system in Region I is singular, and
  reflects the singular nature of the static BTZ coordinate system.}

The diffeomorphism to Kruskal-Szekeres coordinates in Region I can be
obtained by using the definitions \eqref{KruskalUV}:
\begin{align}\label{UVdiffI}
  V &= \eta^{-1} \exp\left(\eta\bar{t} + \eta\int_r^\infty \ud\rho \frac{T(\bar{t}) A(\rho, \bar{t})^{1/2} - \rho}{\rho f(\rho)}\right)\ ,\nonumber\\
    &= \eta^{-1} \exp\left(\eta\bar{t} - \eta\int_r^\infty \ud\rho \frac{\rho A(\rho,\bar{t})}{T(\bar{t}) A(\rho, \bar{t})^{1/2} + \rho}\right)\ ,\nonumber\\
  U &= -\eta^{-1} \exp\left(-\eta\bar{t} - \eta\int_r^\infty \ud\rho \frac{T(\bar{t}) A(\rho, \bar{t})^{1/2} + \rho}{\rho f(\rho)}\right)\ .
\end{align}
The above formulas and can be extended to Region F, $r < R_h$, by
using the Cauchy principal value prescription. In this process, $V$
remains unchanged since the integrand does not have a pole at
$\rho = R_h$ whereas $U$ picks up a minus sign. Thus, in region F, we have
\begin{align}\label{UVdiffF}
  V 
    &= \eta^{-1} \exp\left(\eta\bar{t} - \eta\int_r^\infty \ud\rho \frac{\rho A(\rho,\bar{t})}{T(\bar{t}) A(\rho, \bar{t})^{1/2} + \rho}\right)\ ,\nonumber\\
  U &= \eta^{-1} \exp\left(-\eta\bar{t} - \eta\dashint_r^\infty \ud\rho \frac{T(\bar{t}) A(\rho, \bar{t})^{1/2} + \rho}{\rho f(\rho)}\right)\ .
\end{align}
Similar formulas can be written down for the diffeomorphism of the
fully extended BTZ black hole in Region II and Region P, to the
maximally sliced solution \eqref{maxmetmain}.

\begin{figure}[!ht]
	\centering 
	\includegraphics[width=0.45\linewidth]{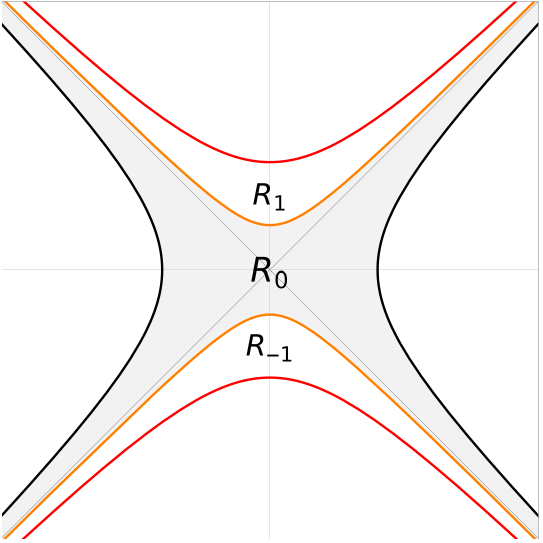}\hfill
	\includegraphics[width=0.45\linewidth]{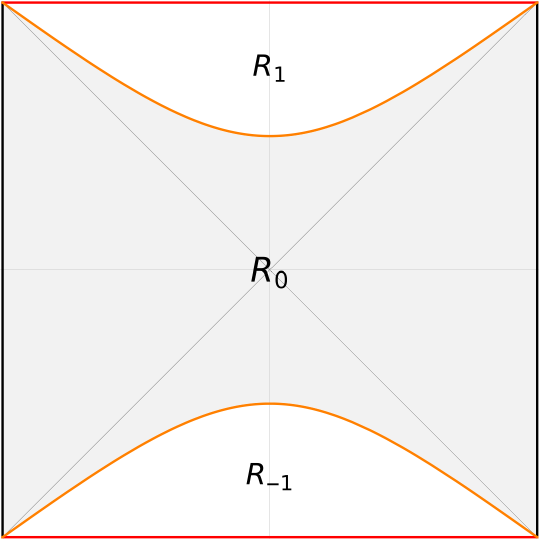}
	\caption{\label{fig:3charts} The region $R_0$ (shaded gray) covered by the
          maximal slicing solution is demarcated by the orange lines
          in the BTZ Kruskal diagram (left) and Penrose diagram
          (right) of the fully extended BTZ black hole. The red curves
          are the singularities at $r=0$ while the black curves are
          the asymptotic boundaries at infinity. The diagonal
          lines are the future and past horizons.}
\end{figure}

\begin{figure}[tbp]
	\centering 
	\includegraphics[width=0.45\linewidth]{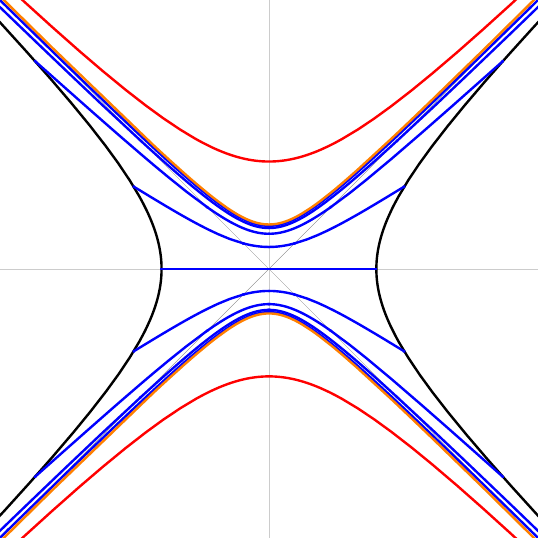}\hfill
	\includegraphics[width=0.45\linewidth]{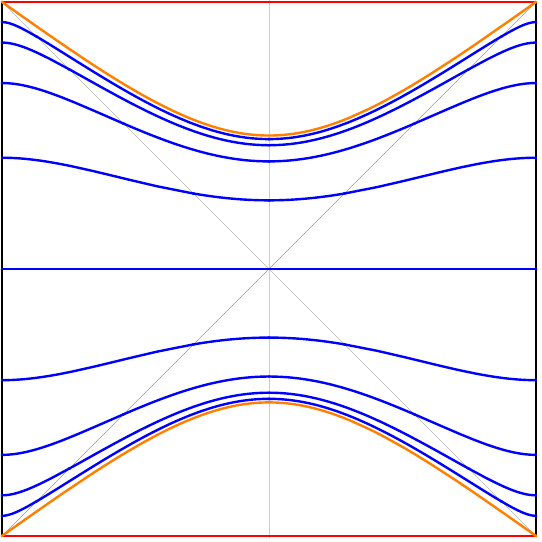}
	\caption{\label{fig:WH1F} Plots of constant $\bar t = 0, \pm 0.5, \pm 1,
          \pm 1.5, \pm 2$ slices, i.e., maximal slices (solid blue lines) in
          the BTZ Kruskal diagram (left) and Penrose diagram (right)
          of the fully extended BTZ black hole. The red curves are the
          singularities at $r=0$ while the black curves are the
          asymptotic boundaries at infinity. As $\bar t\to\infty$, the
          final slice approaches $R_{\infty}=\ell\sqrt{M/4\uppi}$
          (orange curve) from below. Note that the figures are symmetric
          around the reflection slice $\bar t=0$.}
\end{figure}

Finally, by replacing the areal radial coordinates $r$ and $\tl{r}$ by
their expressions in terms of $\theta, \bar{t}$ in $\theta > 0$ and
$\theta < 0$ respectively, we get a diffeomorphism from the spacetime
solution \eqref{ADMmaximal} to the fully extended solution
\eqref{BTZKruskal} in Kruskal-Szekeres coordinates. It is
straightforward to show that this diffeomorphism is completely smooth
everywhere in its domain of definition. However, due to the collapse
of the lapse phenomenon described at the end of Section
\ref{phasetraj}, the maximally sliced solution \eqref{ADMmaximal} is
only mapped to a subregion $R_0$ of the fully extended BTZ black hole
solution. This region $R_0$ is shown in Figure \ref{fig:3charts} and
some representative maximal slices are drawn in Figure \ref{fig:WH1F}.


It is worth noting that the past and future horizons are both covered
by the maximal slicing solution, but their presence is not apparent at
all in the spacetime metric \eqref{ADMmaximal} in the
$(\bar{t}, \theta, \varphi)$ coordinates. The presence and location of
the horizons only become obvious after the diffeomorphism to the
Kruskal-Szekeres coordinates of the fully extended BTZ black
hole.

\subsection{Compatibility with the asymptotic AdS boundary conditions}

The part $\theta > 0$ of the spatial slice $\Sigma$ contains the right
boundary $\theta = + \pi /2$ and the areal radial coordinate $r$ is a
good coordinate in this patch. In the $r \to \infty$ limit, we
approach the right boundary and we check whether the metric, momentum,
lapse and shift written above approach the asymptotic AdS values in
equation \eqref{AdS3split} in Section \ref{AAdSbdry}. In a
neighbourhood of $r \to \infty$, we have the following asymptotic
expressions for the metric,
\begin{align}
  g_{rr} &= \frac{\ell^2}{r^2} \left(1 + \frac{R_h^2}{r^2} + \mc{O}(r^{-4})\right)\ ,\ \  g_{\varphi\varphi} = r^2\ ,\ \ g_{r\varphi} = 0\ ,
\end{align}
for the conjugate momentum,
\begin{align}
  \pi^{rr} &= \frac{T(\bar{t})}{\ell}\left(1 - \frac{R_h^2}{2 r^2} + \mc{O}(r^{-4})\right)\ ,\quad \pi^{r\varphi} = 0\ ,\nonumber\\
  \pi^{\varphi\varphi} &= -\frac{\ell T(\bar{t})}{r^4} \left(1 + \frac{R_h^2}{2 r^2} + \mc{O}(r^{-4})\right)\ ,
\end{align}  
and for the lapse $N$ and shift $N^i$,
\begin{align}
  N(r,\bar{t}) &= \frac{r}{\ell} \left(1 - \frac{R_h^2}{2r^2} + \frac{\ell^3\dot{T}(\bar{t})}{3r^3} + \mc{O}(r^{-4}) \right)\ ,\nonumber\\
N^r(r,\bar{t}) &= \frac{T(\bar{t})}{\ell} \left(1 - \frac{R_h^2}{2r^2} + \frac{\ell^3\dot{T}(\bar{t})}{3r^3} + \mc{O}(r^{-4}) \right)\ ,\nonumber\\
 N^\varphi &= 0\ .
\end{align}
As we can see, the conjugate momentum $\pi^{rr}$ and the shift $N^r$
do not go to zero as $r \to \infty$ but rather to constants which
depend on time, in contrast to the AdS values in
\eqref{AdS3split}. Further, the other components of momentum and shift
fall-off with one power of $r$ smaller than the fall-off conditions
obtained in the classic paper by Brown and Henneaux
\cite{Brown:1986nw}. This is an intriguing aspect of the maximal
slicing gauge, and it seems to require a generalization of the notion
of asymptotic AdS boundary conditions. It is interesting to explore
this further.

\section{Discussion}\label{discsec}

In this paper, we have solved Einstein's equations in the Hamiltonian
formulation in the maximal slicing, spatially harmonic gauge for
spacetimes of the form $\Sigma \times \mathbf{R}$ with spatial slices
$\Sigma$ being cylinders. Before solving the dynamical equations, we
have obtained a reduced phase space for the theory which is finite
dimensional. This reflects the absence of local degrees of freedom in
Einstein's theory of gravitation in $2+1$ dimensions. The phase space
variables $(m, p_m)$ describe global degrees of freedom with the
non-zero number $m$ giving the size of the throat of the cylinder in
the constant curvature metric $\tl{g}_{ij}$. 


Recall that the lapse and shift for the solutions presented in this
paper were solved with the ansatz that they are independent of the
angular coordinate $\varphi$. The most general solution can be written
as a Fourier series in $\varphi$:
\begin{align}
  N(\theta,\varphi) &= A_0(\theta) + \sum_{n = 1}^\infty A_n(\theta) \cos n\varphi + B_n(\theta) \sin n\varphi\ ,\nonumber\\
  N^i(\theta,\varphi) &= C^i_0(\theta) + \sum_{n = 1}^\infty C^i_n(\theta) \cos n\varphi + D^i_n(\theta) \sin n\varphi\ ,
\end{align}
with boundary conditions used in this paper for each of the $A_n$,
$B_n$ and so on. The above represent the most general lapse and shift
with non-trivial boundary conditions and hence the most general large
diffeomorphisms that preserve the gauge conditions. These are expected
to give the generators of the boundary Virasoro algebras (two for each
boundary) discovered by Brown and Henneaux \cite{Brown:1986nw} and the
canonical Poisson brackets of the corresponding Hamiltonians
$H[N, N^i]$ should reproduce the Virasoro algebra with the correct
central charge $c = 3\ell / 2 G_N$.

It is also interesting to explore the origin of the $(m, p_m)$ reduced
phase space in the Chern-Simons formulation of $2+1$ dimensional
Einstein gravity with negative cosmological constant.\footnote{We
  thank O.~Aharony for raising this point.} This is a Chern-Simons
theory with gauge group
$\text{SL}(2,\mathbf{R}) \times \text{SL}(2,\mathbf{R})$, and the $m$
and $T$ reduced phase space variables are most likely related to the
holonomies of the Chern-Simons gauge fields around the
non-contractible cycle of the spatial cylinder.
  
Finally, though we have considered pure gravity in this paper, the
methods used can be extended to gravity coupled to matter fields in
$2+1$ dimensions. In particular, the maximal slicing and spatial
harmonic gauges are still a good set of gauge conditions provided that
scalar fields (if present) have a non-positive potential energy
density. The reduced phase will be infinite dimensional this case due
to the matter degrees of freedom, though there will still be the
finite dimensional sector from the metric fields.

In a companion paper \cite{KPWII}, we consider the simpler problem of a
free massive scalar field probe on the classical maximally sliced
black hole background obtained in this paper. Since the background
solution is completely gauge fixed, there are no residual
diffeomorphisms and the scalar field can be treated as a gauge
invariant object. We quantize the probe scalar field in this maximal
sliced fully extended BTZ black hole background and obtain a unitary
Hamiltonian for the scalar field excitations. The scalar field
correlation functions computed in this maximally sliced background do
not exhibit thermal behaviour, in contrast to the well-known case of
scalar fields in the exterior region of the BTZ black hole.


\begin{acknowledgments}
	We would like to thank Luis Alvarez-Gaume, Suresh Govindarajan, Alok Laddha, Gautam Mandal, Kyriakos Papadodimas and Edward Witten for discussions and comments. 
	We also acknowledge the organizers and participants of `Workshop on Canonical Gravity', held at Chennai Mathematical Institute, for discussions.
	S.~R.~W. would like to thank the CERN Theory Division for
	hospitality, and the Infosys Foundation Homi Bhabha Chair at
	ICTS-TIFR for its support.
\end{acknowledgments}

\appendix

\section{The Lichnerowicz equation}\label{lichapp}

Here we review the proof of the existence and uniqueness of the
solution to the Lichnerowicz solution in $d=2$ using the monotone
iteration method \cite{Andersson:1992yk, Andersson:1996xd,Wittentalk,
  Sakovich:2009nb,Witten:2022xxp}. We only consider pure gravity here,
but the proof can be extended to the case with scalar fields provided
the potential satisfies $V(\phi) \leq 0$. We write the equations in
terms of the extrinsic curvature
\begin{equation}
  \tl{K}^{ij} = \tl{g}^{-1/2} (\tl\pi^{ij} - \tl{g}^{ij} \pi)\ .
\end{equation}
The Lichnerowicz equation is
\begin{equation}\label{d2Licheqapp}
  2 \tl{D}^2\lambda =  \tl{R} - \tl{g}_{ik} \tl{g}_{jl}  \tl{K}^{ij} \tl{K}^{kl}  \e^{-2\lambda} - 2\Lambda \e^{2\lambda}\ .
\end{equation}
Here, $\tl{g}_{ij}$ is a metric with Ricci scalar $\tl{R} =
2\Lambda$. Let us define
\begin{equation}\label{Fdefapp}
  H(\lambda,x) \equiv \tfrac{1}{2}\Big(\tl{R} - \tl{g}_{ik}\tl{g}_{jl} \tl{K}^{ij} \tl{K}^{kl}\e^{-2\lambda} - 2\Lambda \e^{2\lambda}\Big)\ .
\end{equation}
Let $\lambda_\pm$ be a super-solution and sub-solution respectively,
i.e., they satisfy
\begin{equation}
  \tl{D}^2 \lambda_- \geq H(\lambda_-,x)\ ,\quad \tl{D}^2 \lambda_+ \leq H(\lambda_+,x)\ .
\end{equation}
What are appropriate sub- and super-solutions for the current problem?
When $\lambda$ is large and negative, the $\e^{2\lambda}$ in
\eqref{Fdefapp} is negligible and the sum of the first two terms is
large and negative. Thus, $\lambda_- = C_-$ for a large and negative
constant $C_-$ is an appropriate sub-solution. Similarly, when
$\lambda$ is large and positive, the $\e^{2\lambda}$ term dominates if
$V(\phi) \leq 0$; hence, an appropriate super-solution is
$\lambda_+ = C_+$ for a large and positive constant $C_+$. Since the
boundary condition on $\lambda$ is that $\lambda \to 0$ on the
boundaries, we need to take $\lambda = C_\pm$ till some cutoff near
the boundaries and then smoothly, monotonically interpolate to zero as
one approaches either boundary.

Next, define $H_c(\lambda,x) = H(\lambda,x) - c\lambda$. For
a large and positive constant $c$, $H_c(\lambda,x)$ is a
monotonically decreasing function in the interval $[C_-, C_+]$. Define
a sequence of functions $\lambda_0$, $\lambda_1$,\ldots, with
$\lambda_0 = \lambda_+$, and
\begin{equation}
  (\tl{D}^2 - c)\lambda_n = H_c(\lambda_{n-1}, x)\ ,\quad\text{for $n \geq 1$}\ .
\end{equation}
The above equations have unique solutions since $\tl{D}^2 - c$ is
negative definite. Suppose for some $n$, $\lambda_n \leq
\lambda_+$. This certainly true for $n = 0$. Then, we have
\begin{multline}
  (\tl{D}^2 -c)(\lambda_{n+1} - \lambda_+) = H_c(\lambda_{n},x) - \tl{D}^2\lambda_+ + c \lambda_+ \\ \geq H_c(\lambda_n,x) - H(\lambda_+,x) + c\lambda_+\\ = H_c(\lambda_n,x) - H_c(\lambda_+,x) \geq 0\ .
\end{multline}
Suppose the maximum of $\lambda_{n+1} - \lambda_+$ is positive, and is
attained at some point $x_0$. Then, at $x_0$,
$\tl{D}^2(\lambda_{n+1} - \lambda_+)$ is negative, and
$-c(\lambda_{n+1} - \lambda_+)$ is negative as well. Thus,
$ (\tl{D}^2 -c)(\lambda_{n+1} - \lambda_+)$ is negative as well, and
it contradicts the above inequality. Thus, the maximum of
$\lambda_{n+1} - \lambda_+$ is negative and hence
$\lambda_{n+1} \leq \lambda_+$. Thus, by induction, for all $n$, we
have $\lambda_n \leq \lambda_+$.

Similarly, suppose for some $n$, we have $\lambda_- \leq \lambda_n$. Then,
\begin{multline}
  (\tl{D}^2 -c)(\lambda_{-} - \lambda_{n+1}) =  \tl{D}^2\lambda_- - c \lambda_- - H_c(\lambda_{n},x) \\ \geq H(\lambda_-,x) - c\lambda_- - H_c(\lambda_{n},x) \\ = H_c(\lambda_-,x) - H_c(\lambda_{n},x) \geq 0\ .
\end{multline}
If $\lambda_- - \lambda_{n+1}$ has a maximum which is positive, and it
is attained at some point $x'_0$, then
$\tl{D}^2(\lambda_- - \lambda_{n+1}) < 0$ at that point, and so is
$-c(\lambda_- - \lambda_{n+1})$. This contradicts the above inequality
and hence $\lambda_- - \lambda_{n+1} \leq 0$. Thus, we have shown by
induction in $n$ that
\begin{equation}
  \lambda_-\ \leq\ \lambda_n\ \leq\ \lambda_+\ ,\quad\text{for all}\quad n \geq 0\ .
\end{equation}
Next, suppose, for some $n$, $\lambda_{n} \leq \lambda_{n-1}$. This
certainly true for $n = 1$. Then,
\begin{equation}
  (\tl{D}^2-c)(\lambda_{n+1} - \lambda_{n}) = H_c(\lambda_{n},x) - H_c(\lambda_{n-1},x) \geq 0\ .
\end{equation}
If the maximum of $\lambda_{n+1} - \lambda_n$, attained at a point
$x''_0$, is positive, then
$(\tl{D}^2 - c)(\lambda_{n+1} - \lambda_{n}) \leq 0$ at that point
which is in contradiction with the above inequality. Thus,
$\lambda_{n+1} \leq \lambda_n$. By induction, we have proved that
\begin{equation}
  \lambda_{n+1} \leq \lambda_n\ ,\quad\text{for all $n \geq 0$}\ .
\end{equation}
Since the sequence $\{\lambda_n\}_{n \geq 0}$ is bounded and
monotonically decreasing, it has a limit
$$\lim_{n \to \infty} \lambda_n \equiv \lambda\ ,$$ which satisfies the
limit of the iterated equation
\begin{equation}
  (\tl{D}^2 - c)\lambda = H_c(\lambda,x)\ ,
\end{equation}
which implies the Lichnerowicz equation
$\tl{D}^2\lambda = H(\lambda,x)$. Thus, a solution $\lambda$
exists.

To show the uniqueness of the solution, let us first Weyl-transform
back to $g_{ij} = \e^{2\lambda} \tl{g}_{ij}$,
$K^{ij} = \e^{-4\lambda} \tl{K}^{ij}$,
using the solution $\lambda$ which we know exists. Suppose there is
another solution $\lambda' \neq 0$. Then, there is another
configuration $g'_{ij} = \e^{2\lambda'}\tl{g}_{ij}$,
$K'^{ij} = \e^{-4\lambda'} \tl{K}^{ij}$, $\phi' = \phi$,
which satisfies the Hamiltonian constraint. The two solutions are
connected by the Weyl transformation
$\e^{2\lambda''} = \e^{2(\lambda-\lambda')}$ such that
$g'_{ij} = \e^{2\lambda''} g_{ij}$,
$K'^{ij} = \e^{-4\lambda''} K^{ij}$.
The Hamiltonian constraint for
$g'_{ij}$,$K'^{ij}$,
then gives the Lichnerowicz equation for $\lambda''$:
\begin{equation}
  2{D}^2\lambda'' =  {R}  -  {g}_{ik} {g}_{jl}  {K}^{ij} {K}^{kl}  \e^{-2\lambda''} - 2\Lambda \e^{2\lambda''}\ .
\end{equation}
Here $D_i$ is the covariant derivative compatible with $g_{ij}$. Since
$g_{ij}$ and $K^{ij}$ 
satisfy the Hamiltonian constraint, we substitute it into the above
equation to get
\begin{align}
  0 &= 2D^2 \lambda'' +  {g}_{ik}{g}_{jl} {K}^{ij} {K}^{kl}(\e^{-2\lambda''}-1) + 2\Lambda (\e^{2\lambda''}-1)\ \ .
\end{align}
Note that $\Lambda < 0$. If the maximum of $\lambda''$ is strictly
positive, then at the maximum point, we have $D^2\lambda'' \leq 0$, as
well as $\e^{-2\lambda''} > 1$, so that the right hand side is
strictly negative which is a contradiction. Thus, $\lambda'' \leq
0$. Similarly, if the minimum of $\lambda''$ is strictly negative,
then the right hand side is $> 0$ which is again a
contradiction. Thus, $\lambda'' \geq 0$. The only possibility is then
$\lambda'' = 0$ so that $\lambda = \lambda'$ and there is no other
distinct configuration $g'_{ij}$, $K'^{ij}$
that satisfies the Hamiltonian constraint starting from $\tl{g}_{ij}$
and $\tl{K}^{ij}$.

\section{The moduli space of constant negative curvature metrics on
  the cylinder}\label{hyperbolic}

Any two dimensional surface with constant negative curvature can be
obtained as the quotient of the hyperbolic plane by a particular
Fuchsian group, i.e., a discrete subgroup of
$\text{PSL}(2,\mathbf{R})$ (this follows from an application of the
Killing-Hopf theorem on the classification of manifolds with constant
curvature to two dimensional surfaces with constant negative
curvature; see \cite[Section 2]{borthwick_book}, and the recent work
\cite{Chrusciel:2024vle} for a nice review). In particular, for the
cylinder $\mathbf{S}^1 \times \mathbf{R}$, any metric with constant
negative curvature $-2/\ell^2$ is obtained by considering the quotient
of the hyperbolic upper-half plane by a cyclic Fuchsian group
generated by the hyperbolic transformation with length $2\pi |m|$ for
some non-zero real number $m$. By conjugation by an appropriate
$\text{PSL}(2,\mathbf{R})$ element, this hyperbolic transformation can
be taken to be the M\"obius transformation
\begin{equation}
 T = \begin{pmatrix} \e^{\pi |m|} & 0 \\ 0 & \e^{-\pi |m|}\end{pmatrix}\ ,
\end{equation}
and the corresponding Fuchsian group is the cyclic group
$\langle T \rangle$ generated by $T$. The fundamental domain for this
action in the hyperbolic upper-half plane is the strip
$1 \leq |z| < \e^{2\pi |m|}$. Under the identification by the
cyclic group, (1) the two semi-circles are mapped to each other, (2)
the arcs $-\e^{2\pi |m|} < x \leq -1$ and
$1 \leq x < \e^{2\pi |m|}$ on the real line become the boundary
circles, and (3) the segment $1 \leq y < \e^{2\pi |m|}$ on the
positive $y$ axis becomes the closed geodesic of length
$2\pi\ell |m|$ as measured by the hyperbolic metric
$\ell^2 y^{-2}(\ud x^2 + \ud y^2)$.

The induced metric on the cylinder is obtained by first performing the
coordinate transformation $x = \rho \sin\theta$,
$y = \rho \cos\theta$, with
$\theta \in [-\frac{\pi}{2}, \frac{\pi}{2}]$ so that we get
$\ell^2 y^{-2}(\ud x^2 + \ud y^2) = \cos^{-2}\theta (\ud \theta^2 +
\rho^{-2} \ud \rho^2)$. Next, we write $\rho = \e^{{|m|} \varphi}$
so that the metric becomes
\begin{equation}
  \ud\tl{s}^2 = \frac{\ell^2}{\cos^{2}\theta} (\ud \theta^2 + m^2 \ud\varphi^2)\ .
\end{equation}
The identification of $\rho = 1$ and $\rho = \e^{2\uppi |m|}$
implies that $\varphi$ is an angular coordinate with period $2\uppi$.

Thus, the metrics with constant negative curvature on the cylinder are
completely parametrized by the non-zero number $m$.

We can also obtain this result by directly solving the spatial
harmonic gauge conditions and the constant curvature equation
$\tl{R} = 2\Lambda$. It is a well-known result that metric
$\tl{g}_{ij}$ can be written locally (possibly after a diffeomorphism)
as $\tl{g}_{ij} = \e^{2\phi} f_{ij}$ with $f_{ij}$ a constant diagonal
matrix, in which case $\tl{g}_{ij}$ satisfies the spatial harmonic
gauge condition. We make the ansatz that the above decomposition holds
on the entire cylinder $\Sigma$ and that the coordinate system in
which it holds is the $(\theta, \varphi)$ coordinate system described
above. The metric $\tl{g}_{ij}$ is then of the form
\begin{equation}
  \tl{g}_{ij} \ud x^i \ud x^j =  \e^{2\phi} f_{ij}\ud x^i \ud x^j = \e^{2\phi} (a \ud \theta^2 + b \ud\varphi^2)\ ,
\end{equation}
where $a$ and $b$ are positive constants. Plugging the above in to the
constant curvature equation $\tl{R} = 2\Lambda$ and using the fact
that the metric $f_{ij}$ has zero curvature, we get
\begin{equation}\label{liouville}
  2\Lambda = -2\e^{-2\phi}\left(\frac{1}{a} \partial^2_\theta + \frac{1}{b} \partial^2_\varphi\right) \phi\ .
\end{equation}
It is easy to find a solution that is independent of the angular
coordinate $\varphi$. In this case, the equation becomes, using
$\Lambda = -1/\ell^2$,
\begin{equation}
 \frac{\ud^2\phi(\theta)}{\ud\theta^2} = \frac{a}{\ell^2}\e^{2\phi(\theta)}\ .
\end{equation}
It is easy to see that the solution is
\begin{equation}\label{liousol}
  \e^{2\phi(\theta)} = \frac{\ell^2}{a\cos^2\theta}\ .
\end{equation}
It is also easy to see via an application of the maximum principle for
elliptic differential equations that there is no other solution to the
equation \eqref{liouville}. Suppose there is another solution
$\phi'$. Then, there is another metric $g'_{ij} = \e^{2\phi'} f_{ij}$
which satisfies $R' = 2\Lambda$ ($R'$ is the scalar curvature of
$g'_{ij}$). Let $\phi'' = \phi - \phi'$. Then,
$g' = \e^{2\phi''} \tl{g}$, which gives
\begin{equation}
  R' = \e^{-2\phi''} (\tl{R} - 2 \tl{D}^2 \phi'')\ ,
\end{equation}
where $\tl{D}_i$ is the covariant derivative compatible with
$\tl{g}_{ij}$. Plugging in $R' = \tl{R} = 2\Lambda$, we get
\begin{equation}
\tl{D}^2 \phi'' +   \Lambda (\e^{2\phi''} - 1) = 0\ .
\end{equation}
Suppose that the maximum of $\phi''$ is strictly positive. At the
maximum point, we have $\tl{D}^2 \phi'' \leq 0$ and
$ \Lambda (\e^{2\phi''} - 1) < 0$ so that the LHS is strictly negative
which means the equation is not satisfied at such a maximum
point. Thus, the maximum value has to zero or negative and hence
$\phi'' \leq 0$ everywhere. Applying a similar argument to the minimum
value of $\phi''$, we get $\phi'' \geq 0$ everywhere. The only
consistent solution is then $\phi'' = 0$ everywhere, i.e.,
$\phi = \phi'$ everywhere, which shows that the solution in
\eqref{liousol} is unique.

Thus, we get
\begin{equation}
  \tl{g}_{ij} \ud x^i \ud x^j = \frac{\ell^2}{\cos^2\theta} \left(\ud\theta^2 + \frac{b}{a} \ud\varphi^2\right)\ .
\end{equation}
The constant $b/a$ can be identified with the constant $m^2$ of the
previous derivation.

A coordinate system in which the non-compactness of
$\mathbf{R} \times \mathbf{S}^1$ is more manifest is obtained by
setting $\tan \theta = \sinh u$ with $u \in (-\infty, \infty)$ so that
the metric becomes
\begin{equation}
\ud \tl{s}^2 = \ell^2( \ud u^2 + m^2 \cosh^2 u\, \ud \varphi^2)\ .
\end{equation}
A more suggestive coordinate system, which covers only half the
cylinder $u > 0$ or $u \leq 0$, is the areal radial coordinate
$r = \ell {m} \cosh u$ with $r \in (\ell {m}, \infty)$. In
terms of the areal radial coordinate, the metric becomes
\begin{equation}
  \ud \tl{s}^2 = \frac{\ud r^2}{ \frac{r^2}{\ell^2} - m^2} + r^2 \ud \varphi^2\ ,
\end{equation}
which we recognize to be the exterior metric on a constant time slice
of the BTZ black hole \cite{Banados:1992wn,Banados:1992gq}. Starting
from here, one can obtain wormhole-like coordinates which cover the
entire cylinder \cite{Einstein:1935tc, Steif:1995pq, Brill:1995jv} by writing
$w^2 = r^2 - \ell^2 m^2 = \ell^2 m^2 \sinh^2 u$:
\begin{equation}
  \ud \tl{s}^2 =  \ell^2 \left(\frac{\ud w^2}{w^2 + \ell^2 m^2} + (w^2 + \ell^2 m^2 )\ud \varphi^2\right)\ .
\end{equation}

\section{Comparison with Kuchar's phase space for spherically
	symmetric spacetime configurations}\label{Kuchar}

In \cite{Kuchar:1994zk}, Kuchar solved the
Hamiltonian and momentum constraints for the configuration space of
$3+1$ dimensional spherically symmetric asymptotically flat geometries
with two asymptotically flat boundaries. Spherical symmetry implies
that the momentum constraints corresponding to the two sphere
coordinates are automatically satisfied. He then solved the
Hamiltonian constraint and the one remaining momentum constraint to
obtain a finite dimensional reduced phase space.

The same procedure can be repeated for spherically symmetric AdS black
holes in $2+1$ dimensions with asymptotic AdS boundaries with minor
modifications. The result of following Kuchar's procedure of solving
the constraints is again a finite (two) dimensional reduced phase
space parametrized by one coordinate $\bm{M}$ and the conjugate
momentum $\bm{P}$. The Hamiltonian turns out to be
$(c + \tl{c}) \bm{M}$ and it arises from the ADM boundary terms as
obtained previously in Section \ref{redphaseeq}.

The variable $\bm{M}$ has the interpretation that it becomes the ADM
mass of the spacetime solution measured at either of the asymptotic
boundaries. The variable $\bm{P}$ is related to the difference in the
asymptotic AdS time coordinates associated to either AAdS
boundary. These interpretations are borne out of the detailed
expressions for $\bm{M}$ and $\bm{P}$ in Kuchar's analysis; we have
only presented the final results here.

The action in Kuchar's reduced phase space is
\begin{equation}
	\mc{S}[\bm{P},\bm{M}] = \int\ud \bar{t} \Big( \bm{P} \dot{\bm{M}} - (c + \tl{c}) \bm{M}\Big)\ .
\end{equation}
The equations of motion are trivial to solve and we get
\begin{equation}
	\bm{M}(\bar{t}) = \frac{M}{\kappa^2} = \text{constant}\ ,\quad \bm{P}(\bar{t}) = -(c + \tl{c}) \bar{t}\ .
\end{equation}
The obvious question to ask is if there is a classical canonical transformation from the coordinates $(\bm{P}, \bm{M})$ to the coordinates $(Q, P)$ that we have obtained in the present work by complete gauge fixing of the Hamiltonian and momentum constraints. Note that we have also restricted ourselves to the spherically symmetric sector when we chose the solutions of constraints and gauge fixing conditions to be independent of the angular variable $\varphi$.

It turns out that a canonical transformation indeed exists, and in
fact, we already have written the formula for $\bm{M}$ in terms of
$m$, $p_m$ in \eqref{ADMmass}, which is
\begin{equation}
	\bm{M}(Q, P) = \frac{P^2 Q}{2\uppi\ell} \left(k_*\left(\tfrac{Q}{4\uppi^2\ell}\right) + k_*\left(\tfrac{Q}{4\uppi^2\ell}\right)^{-1}\right)\ .
\end{equation}
With some more straightforward manipulations of Kuchar's formulas, the
expression for $\bm{P}$ in terms of $(Q, P)$ can also be obtained:
\begin{equation}
	\bm{P} = \frac{\ell \sgn(Q)}{P}\, \dashint_{-\pi/2}^{\pi/2} \frac{\ud\theta}{ \e^{2\bar\lambda} - (|k_*| + |k_*|^{-1})}\ ,
\end{equation}
where, recall that, $\bar\lambda$ is the solution the Lichnerowicz
equation. Note that there is a simple pole at
$\bar\lambda = \frac{1}{2} \log(k_* + k_*^{-1})$ and the integral has
to be evaluated by the principal value prescription.

Thus, there exists a simpler set of coordinates on the gauge fixed
reduced phase space in which the phase space trajectories are straight
lines parallel to the $\bm{P}$ axis.

We also give a geometrical argument why $(Q, P) \to (\bm{M},\bm{P})$ should be a canonical transformation. The area of an arbitrary region of the phase space is preserved under canonical transformations. We now consider the area below a phase space trajectory, labelled by the mass $M$, in both sets of coordinates and show that they are the same for all phase space trajectories.

The area under a phase space trajectory in terms of $(Q, P)$ coordinates is given by (see figure \ref{phasetrajplot} for reference) 
\begin{equation}
	\frac{1}{\kappa^2} \int \ud Q P(Q) = \frac{1}{\kappa^2} \int \ud \bar{t}\, |\dot{Q}| P(\bar t) = (c+\tilde{c}) \frac{M}{\kappa^2} \int \ud \bar{t} .
\end{equation}
In the second step we switched to time $\bar t$ as the integration variable and in the third step we have used the Hamilton's equation $\dot{Q} = (c+\tilde{c}) M/P$. Similarly, the area under constant $\bm{M}$ straight lines is
\begin{equation}
	\int \ud \bm{P} \bm{M} = \int\ud \bar{t}\, |\dot{\bm{P}}| \bm{M} = (c+\tilde{c})  \int \ud \bar{t}\, \bm{M} ,
\end{equation}
using $\dot{\bm{P}} = -(c+\tilde{c})$ in the last step. Since $\bm{M}=M/\kappa^2$ on-shell, clearly the two areas are equal,
\begin{align}
	\lim_{\mc{T} \to\infty} \frac{\int_{-\mc{T}}^\mc{T} \ud p_m m(p_m)}{\int_{-\mc{T}}^\mc{T} \ud \bm{P}\, \bm{M}} = \lim_{\mc{T} \to\infty} \frac{(c+\tilde{c}) \frac{M}{\kappa^2} \int_{-\mc{T}}^\mc{T}  \ud \bar{t}}{(c+\tilde{c}) \bm{M} \int_{-\mc{T}}^\mc{T}  \ud \bar{t}} =1
\end{align}
for any value of $M$, confirming that $(Q, P) \to (\bm{M}, \bm{P})$ is indeed a canonical transformation.

\section{Diffeomorphism to the Kruskal extension of the BTZ black
  hole}\label{BTZdiffapp}

In this appendix, we show that the maximally sliced spacetime
\eqref{ADMmaximal} obtained by solving the Hamiltonian equations of
motion in the maximal slicing, spatial harmonic gauge is diffeomorphic
to a region $R_0$ of the Kruskal extension of the BTZ black hole. We
reproduce the solution below for convenience:
\begin{align}\label{ADMmaximalapp}
  \ud s^2 &= -N(\bar{t},\theta)^2 \ud \bar{t}^2 + \frac{\ell^2 \e^{2\lambda(\bar{t},\theta)}}{\cos^2\theta} (\ud \theta^2 + N^\theta(\bar{t},\theta)\ud\bar{t})^2 \nonumber\\
  &\qquad\qquad\qquad\qquad\qquad + \frac{\ell^2 \e^{2\lambda(\bar{t},\theta)} }{\cos^2\theta} m(\bar{t})^2 \ud\varphi^2\ .
\end{align}
We first introduce an areal radial coordinate in the solution
\eqref{ADMmaximal}. Using the areal radial coordinate, we describe the
diffeomorphism to the fully extended BTZ black hole.

\subsection{The maximal slicing solution in areal radial
  coordinates}\label{arealsec}

The areal radial coordinate is defined by
$r^2 = g_{\varphi \varphi}(\bar{t}, \theta)$ with $g_{\varphi\varphi}$
obtained from the maximally sliced solution \eqref{ADMmaximalapp}. We
get
\begin{equation}\label{arealrad}
  r(\theta,\bar{t}) = \e^{\lambda(\theta,\bar{t})} \frac{\ell {m(\bar{t})}}{\cos\theta} = \e^{\bar\lambda(\theta,\bar{t})}  \sqrt{\ell|T(\bar{t})|}\ .
\end{equation}
where we have used the definition
$\bar\lambda = \lambda - \log \cos\theta - \frac{1}{2} \log
\frac{|T|}{\ell m^2}$. Recall that $\lambda$ is the unique solution to
the Lichnerowicz equation \eqref{Lichpde}, and it is an even function
of $\theta$. This implies that the above relation is a one-to-one
coordinate transformation either in the range $\theta > 0$ or in the
range $\theta < 0$. The range of $r$ that corresponds to
$0 < \theta < \frac{\pi}{2}$ is
\begin{equation}\label{arealrange-app}
R_+(\bar{t}) = \sqrt{\frac{\ell |T|}{|k_*|}} < r < \infty\ .
\end{equation}
Thus, we need two areal radial coordinates $r$ and $\tl{r}$ with
identical ranges to describe the entire spatial slice:
\begin{align}
  r &= \e^{\lambda(\theta)} \frac{\ell {m}}{\cos\theta} = \e^{\bar\lambda(\theta)}   \sqrt{\ell|T|}\ ,\quad\text{for}\quad \theta > 0\ ,\nonumber\\
  \tl{r} &= \e^{\lambda(\theta)} \frac{\ell {m}}{\cos\theta} = \e^{\bar\lambda(\theta)}  \sqrt{\ell|T|}\ ,\quad\text{for}\quad \theta < 0\ .
\end{align}
In terms of the areal radial coordinate, the metric on $\Sigma$ takes
the form
\begin{equation}
  g_{ij} \ud x^i \ud x^j =  A(r,\bar{t})\ud r^2 + r^2 \ud\varphi^2\ , 
\end{equation}
with
\begin{equation}\label{Artbardef}
  A(r,\bar{t}) = g_{rr} = \left(\frac{r^2}{\ell^2} - \frac{M}{2\uppi} +
  \frac{T(\bar{t})^2}{r^2}\right)^{-1}\ ,
\end{equation}
the extrinsic curvature $K_{ij}$ is
\begin{equation}
  K_{rr} = \frac{T(\bar{t}) A(r,\bar{t})}{r^2}\ ,\quad K_{r\varphi} = 0\ ,\quad  K_{\varphi\varphi} = -T(\bar{t})\ ,
\end{equation}
and the conjugate momentum $\pi^{ij}$ is
\begin{equation}
  \pi^{rr} = \frac{T(\bar{t})}{r A(r,\bar{t})^{1/2}}\ ,\quad  \pi^{r\varphi} = 0\ ,\quad \pi^{\varphi\varphi} = -\frac{T(\bar{t}) A(r,\bar{t})^{1/2}}{r^3}\ .
\end{equation}
There are similar expressions in terms of $\tl{r}$ for the
$\theta < 0$ region of $\Sigma$.

Note that the above expressions are valid only for the range
\eqref{arealrange}. The function $A(r,\bar{t})$ \eqref{Artbardef} has
poles at the roots of the denominator $r = R_\pm (\bar{t})$ which have
the expressions
\begin{equation}
  R_\pm(\bar{t}) = \sqrt{\ell |T| |k_*|^{\mp 1}} = \ell\sqrt{\frac{M}{4\uppi}}\left(1 \pm \sqrt{1 - \frac{16\uppi^2 T^2}{M^2 \ell^2}}\right)^{1/2}\ .
\end{equation}
The parameter $M / 2\uppi$ is the ADM mass of the solution. The roots
satisfy $R_+(\bar{t}) \geq R_-(\bar{t})$ with equality satisfied at
$\bar{t} \to \infty$. Hence, the second root $R_-(\bar{t})$ lies
outside the range \eqref{arealrange}.

The lapse is given by
\begin{align}
  N(r,\bar{t}) &=   \sqrt{\frac{r^2}{\ell^2} - \frac{M}{2\uppi} + \frac{T(\bar{t})^2}{r^2}} \times\nonumber\\
  &\qquad \times \left(1 + \dot{T}(\bar{t}) \int_\infty^r \frac{\ud \rho}{\rho\left(\frac{\rho^2}{\ell^2} - \frac{M}{2\uppi} + \frac{T(\bar{t})^2}{\rho^2}\right)^{3/2}} \right)\ ,
\end{align}
which is valid for the range \eqref{arealrange} and the same
expression in terms of $\tl{r}$ for $\theta < 0$.


The non-trivial component of the shift is $N^\theta$ which can be
transformed to areal radial coordinates according to the following rule:
\begin{equation}
  \frac{\partial r}{\partial \theta} N^\theta = \frac{\partial r}{\partial \bar{t}} + N^r\ .
\end{equation}
The term $\partial r / \partial \bar{t}$ on the right hand side is
present to account for the change in the relation of the areal radial
coordinate $r$ to $\theta$ under time evolution. The shift in areal
radial coordinates comes out to be
\begin{equation}
  N^r = \frac{N T}{r}\ ,\quad N^\varphi = 0\ .
\end{equation}
The spacetime metric corresponding to a given trajectory in reduced
phase space is given by
\begin{align}\label{maxmet}
  &{}^{(2+1)}\ud s^2 =\nonumber\\
  & -N^2(r, \bar{t}) \ud \bar{t}^2 + A(r,\bar{t}) \left( \ud r + \frac{N(r,\bar{t}) T(\bar{t})}{r} \ud \bar{t}\right)^2 + r^2 \ud\varphi^2\ .
\end{align}
The collapse of the lapse that was described in \eqref{collapse} of
the previous section implies that the above solution approaches a
limit surface
$r_{\rm min} = \lim _{\bar {t} \to \infty} R_+(\bar{t}) = \ell \sqrt{M
  / 4\uppi}$ as $\bar{t} \to \infty$. In particular, the local spatial
volume governed by $\sqrt{g} = r\sqrt{A}$ never goes to zero since $r$
never reaches $0$. This behaviour is typical of maximal slices since
they tend to maximize the local volume, whereas going to smaller and
smaller values tends to minimize the local volume. In higher
dimensions, where $r = 0$ corresponds to a genuine curvature
singularity of the volume-crushing type (e.g.the ~AdS-Schwarzschild
solution in $3+1$ dimensions), the maximally sliced solution with
wormhole spatial slices avoids the singularity altogether.

\subsection{The diffeomorphism}\label{BTZdiff}

The final metric \eqref{maxmet} is written in terms of an areal radial
coordinate. This facilitates an easy map to the BTZ black hole metric
in Region I which is also in terms of an areal radial coordinate:
\begin{equation}\label{BTZmet}
  \ud s^2 = - f(r) \ud t^2 + \frac{\ud r^2}{f(r)} + r^2 \ud\varphi^2\ ,\ \ f(r) = \frac{r^2}{\ell^2} - \frac{M}{2\uppi}\ .
\end{equation}
The diffeomorphism from Region I of the BTZ black hole to the
maximally sliced solution is then
\begin{equation}
(r,\varphi, t) = \big(r, \varphi, t = t(\bar{t}, r)\big)\ .
\end{equation}
Plugging $t = t(\bar{t}, r)$ into the BTZ metric \eqref{BTZmetmain},
and comparing with the maximal slicing metric \eqref{maxmet}. We get the
equations
\begin{equation}
  \left(\frac{\partial t}{\partial \bar{t}}\right)^2 = N(r, \bar{t})^2 A(r, \bar{t})\ ,\quad   \left(\frac{\partial t}{\partial r}\right)^2 = \frac{T(\bar{t})^2 A(r,\bar{t})}{r^2 f(r)^2}\ .
\end{equation}
In taking the square roots, we have a choice of a sign. The signs in
Region I have to be chosen such that increasing $t$ corresponds to
increasing $\bar{t}$ (with $r$ fixed) and decreasing $r$ (with
$\bar{t}$ fixed). This gives, in Region I with $\bar{t} > 0$,
\begin{equation}
  \frac{\partial t}{\partial \bar{t}} = N(r, \bar{t}) A(r, \bar{t})^{1/2}\ ,\quad   \frac{\partial t}{\partial r} = -\frac{T(\bar{t}) A(r,\bar{t})^{1/2}}{r f(r)}\ .
\end{equation}
As $r \to \infty$, we have $N \sqrt{A} \to 1$. This allows the boundary condition
\begin{equation}
  \lim_{r \to \infty} t(r,\bar{t}) = \bar{t} + t_0\ ,
\end{equation}
for some arbitrary constant $t_0$. We typically choose $t_0 = 0$ so
that the origin of time in $t$ and $\bar{t}$ coincide. The
diffeomorphism in Region I is given by integrating the second equation
at constant $\bar{t}$:
  \begin{equation}
  t(r,\bar{t}) = \bar{t} + T(\bar{t}) \int_r^\infty \frac{\ud \rho}{\rho} \frac{A(\rho,\bar{t})^{1/2}}{f(\rho)}\ ,\quad r > R_h\ .
\end{equation}
It is easy to see that differentiating the above solution at constant
$\bar{t}$ reproduces the expression for
$N(r,\bar{t}) A(r,\bar{t})^{1/2}$.

\bibliography{refs}

\end{document}